\documentclass{article}
\usepackage{arxiv}
\usepackage[utf8]{inputenc} 
\usepackage[T1]{fontenc}    
\usepackage{setspace}
\usepackage{hyperref}       
\usepackage{url}            
\usepackage{booktabs}       
\usepackage{amsfonts}       
\usepackage{nicefrac}       
\usepackage{microtype}      
\usepackage{graphicx}
\usepackage{natbib}
\usepackage{lineno}
\usepackage{dsfont}
\usepackage{doi}
\usepackage{amsthm,amsmath,amssymb}
\usepackage{graphicx,bm}
\usepackage{color}
\usepackage{lscape}
\usepackage{rotating}
\usepackage{setspace}
\usepackage{algorithm}
\usepackage{algpseudocode}
\usepackage{epigraph} 
\usepackage{subfigure}

\title{Reduced Rank Regression for Mixed Predictor and Response Variables}

\author{ \href{https://orcid.org/0000-0001-7308-6210}{\includegraphics[scale=0.06]{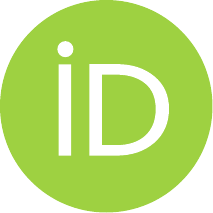}\hspace{1mm}Mark de Rooij}\\
	Methodology and Statistics department\\
    Leiden University\\
	The Netherlands \\
	\texttt{rooijm@fsw.leidenuniv.nl} \\
	\And
	\href{https://orcid.org/0009-0008-6752-1392}{\includegraphics[scale=0.06]{orcid.pdf}\hspace{1mm}Lorenza Cotugno} \\
	Department of Physics\\
   University of Naples Federico II\\
	Italy \\
	\texttt{lorenza.cotugno@unina.it} \\	
	\And
	\href{https://orcid.org/0000-0002-8012-0461}{\includegraphics[scale=0.06]{orcid.pdf}\hspace{1mm}Roberta Siciliano} \\
    Department of Electrical Engineering and Information Technology \\
    University of Naples Federico II\\
	Italy \\
	\texttt{roberta.siciliano@unina.it} \\	
}

\date{}

\renewcommand{\headeright}{}
\renewcommand{\undertitle}{}
\renewcommand{\shorttitle}{GMR\textsuperscript{3}}

\begin{document}
\maketitle

\begin{abstract}
In this paper, we propose the generalized mixed reduced rank regression method, GMR$^3$ for short. GMR$^3$ is a regression method for a mix of numeric, binary, and ordinal response variables. The predictor variables can be a mix of binary, nominal, ordinal, and numeric variables. For dealing with the categorical predictors we use optimal scaling. A majorization-minimization algorithm is derived for maximum likelihood estimation under a local independence assumption. A series of simulation studies is shown (Section \ref{sec:simstudy}) to evaluate the performance of the algorithm with different types of predictor and response variables. In Section \ref{sec:modelselection}, we briefly discuss the choices to make when applying the model the empirical data and give suggestions for supporting such choices. In Section \ref{sec:data}, we show an application of GMR$^3$ using the Eurobarometer Surveys data set of 2023.
\end{abstract}


\section{Introduction}

In this paper, we will describe a general methodology for the case where we are interested in the analysis of dependence, that is, we like to see the dependence of a set of variables on another set of variables, as in regression methods. Such an analysis of dependence can be contrasted to an analysis of interdependence \citep{gifi1990nonlinear}, the latter treats the variables in a symmetric way, as in correlation and association analysis. 

Usually, researchers have multiple response variables and multiple predictor variables. Typically, researchers analyze the data in a univariate way, one response variable at a time. However, as \cite{fish1988multivariate} argued, it is important to analyze the data using multivariate methods, as they best honor the reality about which the researcher is purportedly trying to generalize. To quote \cite{fish1988multivariate}, "\emph{The reality in which social scientists are interested is usually one in which the researcher cares about multiple outcomes, in which most outcomes have multiple causes, and in which most causes
have multiple effects}". 

Variables can be measured on different scales \citep{stevens1946theory}. It is possible to distinguish between ratio, interval, ordinal, and nominal variables. Binary variables can be considered special cases of these. In a statistical analysis, it is important to take into account the measurement level of the variable, as this defines the kind of operations that are allowed. In the multivariate social context defined above, both the responses and the predictors may have varying measurement levels.  

In this paper, we consider numeric (i.e., ratio and interval responses), binary, and ordinal response variables. These types of response variables are typical for social science applications. For numerical responses usually a linear regression model is used, where the expected value of the response variable is linked to a linear combination of the predictor variables. For binary responses the expected value is a probability, which is linked to the linear predictor by the logistic link function. Finally, for ordinal response variables, researchers typically use the cumulative logistic regression model \citep{mccullagh1980regression, agresti2013categorical}. One interpretation of this ordinal regression model is based on an underlying continuous latent variable that is partitioned in a set of ordered classes using a set of thresholds or cutpoints. 

On the predictor side of the model, we consider variables that might be numeric, ordinal, nominal, or binary. To take the measurement level into account, optimal scaling can be employed \citep{gifi1990nonlinear}. In optimal scaling, a categorical variable is replaced by a set of quantifications, that are transformations (i.e., scaling) of a variable $\bm{\phi}_p = \varphi_p(\bm{x}_p)$. The transformation functions, $\varphi_p(\cdot)$, are variable specific and take into account the measurement level of the variable. The result of the transformation of variable $\bm{x}_p$ is a new, optimally scaled, variable  $\bm{\phi}_p$, optimal in the sense that the transformation minimizes a loss function. 
Regression with optimal scaling was first proposed by \cite{young1976regression}. A detailed treatment was recently given by \cite{meulman2019ros}. \cite{willems2020advances} proposed to use optimal scaling of categorical predictors in generalized linear models and survival analysis.

For the analysis of dependence with multiple numeric predictor and multiple numeric response variables, a multivariate regression model can be used. In such a multivariate model, for every predictor-response pair a regression weight is estimated. To reduce the number of parameters, the matrix of regression weights might be constrained to be a product of two thin matrices. In such a way, a rank constraint is imposed on the matrix of regression weights. This type of model was first introduced by \cite{anderson1951estimating} and further developed by several authors \citep{izenman1975reduced, tso1981reduced, davies1982procedures, wollenberg1977redundancy}. The reduced rank regression model for numeric response variables, $r = 1,\ldots, R$ can be written as
\[
y_{ir} = m_r + \bm{x}_i'\bm{B}\bm{v}_r + \epsilon_{ir}, \ \forall \ r 
\]
where $m_r$ is a conditional mean, or intercept for response variable $r$, $\bm{x}_i$ is the vector with observed values of the predictor variables for observation $i$, $\bm{B}$ is a matrix with regression coefficients, $\bm{v}_r$ is a vector of loadings for response variable $r$ and $\epsilon_{ir}$ is the  error component. More detailed information about this model will follow in the next section. Reduced rank regression, or redundancy analysis, can also be considered a constraint principal component analysis \citep{takane2013constrained}.  In that case, the object scores of principal component analysis are constrained to be a linear combination of the predictor variables. The reduced rank model has been generalized to other types of response variables. \cite{yee2015} extends the models for response variables in the exponential family, whereas \cite{derooij2023new} developed an extension specifically for binary response variables. For ordinal variables, \cite{derooijbreemerwoestenburgbusing2022}, recently described a reduced rank model with a cumulative link function. 

These reduced rank models consider one type of response variables, that is, all responses should be numeric, or all should be binary, or all should be ordinal. In empirical research, however, typically the responses have different measurement scales. In the example considered in Section \ref{sec:applications}, for example, there are binary and ordinal response variables.  \cite{luo2018leveraging} developed a reduced rank model for different types of responses. They did consider numeric, binary and count variables, but not ordinal ones. In the social sciences, ordinal response variables are most common. Furthermore, the different measurement levels on the predictor side are usually taken into account by creating a set of dummy variables. Specifically, for ordinal variables this approach does not take into account the nature of the variable. We will include optimal scaling of the predictor variables.  

In this manuscript, we will develop a reduced rank regression model for mixed numeric, ordinal, and binary response variables with optimal scaling for the predictor variables. In Section \ref{sec:model}, we present the model and its interpretation. In Section \ref{sec:algo}, we present a majorization-minimization (MM) algorithm for maximum likelihood estimation of the parameters of the model. In Section \ref{sec:modsel}, we describe model selection issues in detail. 
In Section \ref{sec:applications}, we will describe an application of the methodology to survey data obtained through the Eurobarometer studies, where the responses are a mixture of binary and ordinal variables, and the predictors a mixture of numerical, nominal, and ordinal variables. We end this paper with a general discussion and conclusion.

\section{Generalized Mixed Reduced Rank Regression}\label{sec:model}

We will have a set of $P$ predictor and 
$R$ response variables. The response variables are indexed by $r$ ($r = 1\ldots, R$), while the predictors are indexed by $p$ ($p = 1\ldots, P$). The data are collected for $N$ participants. The observations for participant $i$ ($i = 1\ldots, N$) are denoted by $\bm{y}_i$ and $\bm{x}_i$, for the responses and predictors, respectively. 

In what follows, we first describe the treatment of predictor variables after which we describe how we treat the different types of response variables. We end this section with the likelihood equations for our model. 

\subsection{Predictor Variables}

We have $P$ predictor variables, that are partitioned in a numeric set ($\mathcal{N}_p$), and another set with discrete predictor variables, that are nominal, ordinal and binary predictor variables ($\mathcal{D}_p$). For the numeric variables, we quantify them by standardizing the values, that is 
\[
\phi_p = \frac{x_p - \bar{x}_p}{\mathrm{sd}(x_p)},
\]
where $\mathrm{sd}()$ computes the standard deviation. 

For the discrete (non-numeric) predictor variables, indicator matrices $\bm{G}_p$ of size $N \times C_p$, where $C_p$ is the number of categories of predictor $p$, are defined. An optimally scaled variable is obtained by 
\[
\varphi_p(\bm{x}_p) = \bm{G}_p\bm{w}_p
\]
where $\bm{w}_p$ are quantifications, to be estimated. The quantifications are estimated such that they minimize the loss function (i.e., the negative log likelihood, see next Section). The $\bm{w}_p$ obtained in this way, are optimal for binary and nominal predictor variables. For ordinal predictor variables an extra step is needed, as the quantifications may not be ordered correctly. Therefore, the unconstrained quantifications are projected on the cone of admissable transformations. For an ordinal scaling level, this amounts to performing a monotone regression \citep{deleeuw2005monotonic, busing2022monotone}. 

The transformed predictor variables will be collected in the $N \times P$ matrix $\bm{\Phi}$, 
\[
\bm{\Phi} = \left[ \bm{\phi}_1, \ldots, \bm{\phi}_P \right] = \left[ \varphi_1(\bm{x}_1), \ldots, \varphi_P(\bm{x}_P) \right].
\]
The elements of a row of the matrix $\bm{\Phi}$ will be collected in the $P$-dimensional column vector $\bm{\phi}_i$, representing the optimally transformed variables for observation $i$ ($i = 1,\ldots, N)$. 

\subsection{Response Variables}

The $R$ response variables ($r = 1\ldots, R$) are partitioned in three sets: 
a set of numeric variables $\mathcal{N}$, 
a set of binary variables $\mathcal{B}$, 
and a set of ordinal variables $\mathcal{O}$. 

We define the canonical term $\theta_{ir}$ and the following bilinear  or reduced rank structure is imposed
\[
\theta_{ir} = m_r + \bm{\phi}_i'\bm{B}\bm{v}_r, 
\]
where $m_r$ is an intercept, $\bm{\phi}_i$ are the optimally scaled predictor values for participant $i$, $\bm{B}$ are regression weights to be estimated and $\bm{v}_r$ are loadings for the $r$-th response variable. The matrix $\bm{B}$ is of size $P \times S$ and the vector of loadings for response variable $r$ has length $S$. This number has to be chosen by the researcher, it is the required rank or \emph{dimensionality}. The loadings $\bm{v}_r$ can be collected in the $R \times S$ matrix $\bm{V}$ as
\[
\bm{V} = \left[\bm{v}_1, \ldots, \bm{v}_r, \ldots, \bm{v}_R \right]'.
\]
For identification, we require the matrix $\bm{V}$ to be orthogonal, that is $\bm{V}'\bm{V} = \bm{I}$, where $\bm{I}$ is the identity matrix of order $S$. Furthermore, we require $\bm{U}'\bm{U}$ to be a diagonal matrix, where $\bm{U} = \bm{\Phi}\bm{B}$. 

For numeric responses, $\theta_{ir}$ represents the expected or estimated value of the response $r$. We assume a normal distribution for the numeric response variables.

Similarly, for binary response variables, $\theta_{ir}$ represent the log-odds form, that is, 
\[
\log \frac{\pi_{ir}}{1 - \pi_{ir}} = \theta_{ir}.
\] 
We assume the binary response variables to have a Bernoulli distribution with probability $pi_{ir}$.

For ordinal variables, the story is a bit different. The number of categories of response variable $r$ is $C_r$, coded as $c = 1, \ldots, C_r$. Underlying each ordered categorical response variable $y_r$ we assume a continuous latent variable $y^*_r$. 
These latent variables are modeled as
\[
y^*_{ir} = \theta_{ir} + \epsilon_{ir},
\]
where $\theta_{ir}$, the canonical parameter is defined as
\[
\theta_{ir} = \bm{\phi}_i'\bm{B}\bm{v}_r.
\]
We see that for ordinal response variables the $m_r = 0$, because, without loss of generality, we can assume that the latent underlying continuous response variable is centered, so no intercept is needed. The continuous underlying variable is partitioned through a set of cut-points or \emph{thresholds} to form a set of ordered categories. Let $-\infty = t_0 < t_1<  \ldots < t_{C_r} = \infty$ define the set of thresholds such that an observed ordinal response $y_r$ satisfies
\[
y_r = c \ \  \mathrm{if} \ t_{c-1} \leq y_r^* < t_c, 
\]
for $c = 1, \ldots, C_r$.  It is typically assumed that the $\epsilon_{ir}$ are independent and identically distributed error terms following a cumulative logistic distribution $F(\cdot)$, that is,
\[
F(\eta) = \frac{1}{1 + \exp(-\eta)} \ \mathrm{for} \ \eta \in \ (-\infty, \infty).
\]
It follows that 
\[
F^{-1}\left(P(y_{ir} \leq c) \right) = \log \left( \frac{P(y_{ir} \leq c)}{P(y_{ir} > c)} \right) = t_{r_c} -  \theta_{ir},
\]
where, similar to the proportional odds regression model \citep{mccullagh1980regression, anderson1981regression}, the thresholds are \emph{category specific} but the structural part of the model is \emph{variable specific}. From this specification of the model for ordinal variables, it follows that the probability that person $i$ will respond with category $c$ on response variable $r$ is 
\[
\pi_{irc} = P(y_{ir} \leq c) - P(y_{ir} \leq c - 1), \ \mathrm{for} \ r \in \mathcal{O},
\]
a probability that we need to define the log-likelihood function. The ordered response follows a multinomial distribution with probabilities $\pi_{irc}$.

Summarizing, for numeric, binary, and ordinal response variables the canonical parameters are defined in terms of a conditional mean or intercept ($\bm{m}$), the regression weights ($\bm{B}$), and the loadings ($\bm{V}$). The intercepts for ordinal variables are, by definition, equal to zero. Instead, for these ordinal variables thresholds are defined ($\bm{t}$) that partition the latent underlying continuous variable into a ordinal categorical observed variable. 

\subsection{Log-likelihood function}

The parameters of the generalized mixed reduced rank models are estimated by maximum likelihood. We will denote the negative of the log-likelihood function by $\mathcal{L}(\bm{\theta}, \bm{t})$. The canonical parameters ($\bm{\theta}$) , will be later parameterized by the intercepts ($\bm{m}$), weights ($\bm{B}$), loadings ($\bm{V}$), and the quantifications ($\bm{w}$). The vector $\bm{t}$ collects the set of threshold parameters for ordinal response variables.

We assume conditional independence, that is, given the low dimensional embedding, the response variables are independent, such that the negative log likelihood partitions in contributions of the single response variables, that is, 
\[
\mathcal{L}(\bm{\theta}, \bm{t}) = \sum_{r} \mathcal{L}_r(\bm{\theta}, \bm{t}), 
\]
where $\mathcal{L}_r(\bm{\theta}, \bm{t})$ depends on the set of response variable $r$, that is
\[
\mathcal{L}_r(\bm{\theta}, \bm{t}) = \left\{ 
\begin{array}{lr} 
\sum_i \frac{1}{2\sigma^2} (y_{ir} - \theta_{ir})^2 + N \log(\sqrt{2 \pi \sigma^2})  & \mathrm{if}\  r \in \mathcal{N} \\
\sum_i -\log((1 + \exp(-q_{ir}\theta_{ir})^{-1})) & \mathrm{if}\  r \in \mathcal{B} \\
\sum_i \sum_{c} - g_{irc}\log \pi_{irc} & \mathrm{if}\  r \in \mathcal{O} 
\end{array} 
\right.,
\]
where the expression for the binary variables was derived in \cite{deleeuw2006principal} with $q_{ir} = 2y_{ir} - 1$. 

\section{Algorithm}\label{sec:algo}

The general algorithm will alternate between updating the canonical part and the threshold parameters. When updating the canonical parameters, we assume that the threshold parameters are fixed, and \emph{vice versa}. To avoid cluttering of notation, we will write 
$\mathcal{L}(\bm{\theta}) = \mathcal{L}(\bm{\theta}, \bm{t})$ when we update the canonical parameter 
and 
$\mathcal{L}(\bm{t}) = \mathcal{L}(\bm{\theta}, \bm{t})$ when updating the thresholds. 

To update the canonical part, we employ an MM algorithm, where MM stands for Majorization Minimization \citep{heiser1995convergent, hunter2004tutorial, nguyen2017introduction}. The concept behind MM, applied to finding a minimum of the function \(\mathcal{L}(\bm{\theta})\), with
\(\bm{\theta}\) representing a vector of parameters, involves defining an auxiliary function known as a  \emph{majorization function}, 
\(\mathcal{M}(\bm{\theta}|\bm{\vartheta})\). The vector $\bm{\vartheta}$ is a vector with so-called support points of the same length as the vector $\bm{\theta}$. In the iterative algorithm, the support points are usually given by the values of the parameters at that stage in the algorithm. The majorization function has  two key characteristics
\[
\mathcal{L}(\bm{\vartheta}) = \mathcal{M}(\bm{\vartheta}|\bm{\vartheta})\\
\] 
and
\[
\mathcal{L}(\bm{\theta}) \leq \mathcal{M}(\bm{\theta}|\bm{\vartheta}).
\] 
These equations indicate that \(\mathcal{M}(\bm{\theta}|\bm{\vartheta})\) is a function positioned above (i.e., majorizes) the original function, touching it at the supporting point. An iterative algorithm can be formulated as
\[
\mathcal{L}(\bm{\theta}^+) \leq \mathcal{M}(\bm{\theta}^+|\bm{\vartheta}) \leq \mathcal{M}(\bm{\vartheta}|\bm{\vartheta}) = \mathcal{L}(\bm{\vartheta}),
\] 
where \(\bm{\theta}^+\) is determined as
\[
\bm{\theta}^+ = \mathrm{argmin}_{\bm{\theta}} \ \mathcal{M}(\bm{\theta}|\bm{\vartheta}) \, ,
\] 
representing the updated parameter, which becomes the $\bm{\vartheta}$ in the next iteration.  

MM algorithms have several properties. The first is that it usually has a simple numerical minimization method. We will see shortly that in our case the majorization function is a least-squares problem that is relatively easy to solve. Furthermore, the value of the loss function (i.e., in our case the negative log-likelihood) should never increase, which makes it easy to check the programming. MM algorithms are globally convergent and usually end in a local minimum. The disadvantages of MM algorithms are that they are slow (linear convergence rate) and it may be difficult to prove convergence of the parameters.

\subsection{MM algorithm for canonical part}\label{sec:mm}

The negative log likelihood for response variable $r$ is defined as a sum over individual parts, that is
\[
\mathcal{L}_r(\bm{\theta}) = \sum_{i=1}^N \mathcal{L}_{ir}(\theta_{ir}).
\]
Finding a majorization function for each $\mathcal{L}_{ir}(\theta_{ir})$ also gives a majorization function for the sum. Looking at a single element, where we omit the subscripts for the moment, the \emph{quadratric majorization theorem} states that
\[
\mathcal{L}(\theta) \leq \mathcal{L}(\vartheta) + \frac{\partial \mathcal{L}(\theta)}{\partial \theta} (\theta - \vartheta) + \frac{1}{2}(\theta - \vartheta)\kappa(\theta - \vartheta) = \mathcal{M}(\theta| \vartheta)
\]
for any $\kappa \geq \psi = \frac{\partial^2 \mathcal{L}(\theta)}{\partial \theta^2}$. 

Denote \(\xi = \frac{\partial \mathcal{L}(\theta)}{\partial \theta}\), and let us rewrite step by step the majorization function $\mathcal{M}(\theta| \vartheta)$, that is
\[
\begin{aligned}
\mathcal{M}(\theta, \vartheta) &= 
\mathcal{L}(\vartheta) + \xi (\theta - \vartheta) + \frac{1}{2}(\theta -\vartheta)\kappa(\theta - \vartheta) \\
&= \mathcal{L}(\vartheta) + \xi\theta - \xi\vartheta + \frac{\kappa}{2}(\theta^2 + \vartheta^2 - 2\theta\vartheta) \\
&= \frac{\kappa}{2}\theta^2 + 2\frac{\kappa}{2}\theta\left(\frac{1}{\kappa}\xi - \vartheta\right) + c_1 \\
&= \frac{\kappa}{2}\theta^2 - 2\frac{\kappa}{2}\theta\left(\vartheta - \frac{1}{\kappa}\xi\right) + c_1 \\
&= \frac{\kappa}{2}\left(\theta^2 - 2z\theta\right) + c_1 \\
&= \frac{\kappa}{2}\left(\theta^2 - 2z\theta + z^2\right) -\frac{\kappa}{2} z^2 + c_1 \\
&= \frac{\kappa}{2}\left(\theta - z\right)^2 -\frac{\kappa}{2} z^2 + c_1 \\
&= \frac{\kappa}{2}\left(\theta - z\right)^2  + c \\
\end{aligned}
\]
where $z = (\vartheta - \frac{1}{\kappa}\xi)$, a working response and $c = c_1 -\frac{\kappa}{2} z^2$ and $c_1 = \mathcal{L}(\vartheta) - \xi\vartheta + \frac{\kappa}{2}\vartheta^2$, all constants with respect to $\theta$. The last line shows that the majorization function is a least-squares function. For the different types of response variables, we have to derive the expression of the first derivatives \(\xi\) and the majorizing constant \(\kappa\).

\paragraph{Majorization function for numeric response variables:}

The loss function is
\[
\mathcal{L}_{ir}(\theta_{ir}) = \frac{1}{2\sigma^2} (y_{ir} - \theta_{ir})^2 + N \log(\sqrt{2 \pi \sigma^2}).
\]
The first derivative of \(\mathcal{L}_{ir}(\theta_{ir})\) with respect to
\(\theta_{ir}\) is \[
\begin{aligned}
\xi_{ir} \equiv \frac{\partial \mathcal{L}_{ir}(\theta_{ir})}{\partial \theta_{ir}} = \frac{1}{\sigma^2}(\theta_{ir} - y_{ir})
\end{aligned}
\]
and the second derivative is 
\[
\psi_{ir} \equiv \frac{\partial^2 \mathcal{L}_{ir}(\theta_{ir})}{\partial \theta_{ir}^2} = \frac{1}{\sigma^2} 
\]
so that an upper bound is obtained for any $\kappa \geq  \sigma^{-2}$. This majorization function was also used by \cite{song2021generalized} in their approach for principal component analysis of binary and numeric variables.

\paragraph{Majorization function for binary response variables:}

The loss function is
\[
\mathcal{L}_{ir}(\theta_{ir}) = -\log \frac{1}{1 + \exp(-q_{ir}\theta_{ir})}.
\]
The first derivative of \(\mathcal{L}_{ir}(\theta_{ir})\) with respect to
\(\theta_{ir}\) is \[
\begin{aligned}
\xi_{ir} \equiv \frac{\partial \mathcal{L}_{ir}(\theta_{ir})}{\partial \theta_{ir}} & = -(y_{ir} - \pi_{ir})
\end{aligned}
\] 
and the second derivative is 
\[
\psi_{ir} \equiv \frac{\partial^2 \mathcal{L}_{ir}(\theta_{ir})}{\partial \theta_{ir}^2} = \pi_{ir}(1 - \pi_{ir}) 
\]
so that an upper bound is obtained for any $\kappa \geq \frac{1}{4}$.
This majorization function was derived by \cite{deleeuw2006principal}. 

\paragraph{Majorization function for ordinal response variables:}

For ordinal variables, we start out a bit different because we deal with a latent variable and therefore aim for an EM algorithm. In the EM algorithm, the first step is to define the \emph{complete data negative log-likelihood}, that is, the likelihood assuming that we have observed the underlying latent variable. In the E-step, the expected value of this complete data negative log-likelihood is obtained, which in the M-step is minimized. For minimization, we use an upper bound again like in the MM algorithm. 

An element of the complete data negative log-likelihood is 
\[
\mathcal{L}^c_{ir}(\theta_{ir}) = - \log f(y^*_{ir} - \theta_{ir}), 
\]
where \(f(\cdot)\) is the probability density function of the logistic distribution. The expected value of the 
second-order Taylor expansion of the complete data negative log-likelihood around the current value $\vartheta$ is 

\[
\mathbb{E}(\mathcal{L}^c_r(\theta_{ir})) = 
\mathbb{E}(\mathcal{L}^c_r(\vartheta_{ir})) + (\theta_{ir} - \vartheta_{ir}) \mathbb{E}\left(\frac{\partial \mathcal{L}^c_r(\vartheta_{ir})}{\partial \theta_{ir}}\right)
+ \frac{1}{2} (\theta_{ir} - \vartheta_{ir}) \mathbb{E}\left(\frac{\partial^2 \mathcal{L}^c_r(\vartheta_{ir})}{\partial^2 \theta_{ir}}\right) (\theta_{ir} - \vartheta_{ir}).
\]
Let us define \( p_{ir} = \frac{1}{1 + \exp(-y^*_{ir} + \theta_{ir})}\) so that \( \log f(y^*_{ir} - \theta_{ir}) = \log p_{ir}(1 - p_{ir})\). The partial derivative is
\[
\frac{\partial \mathcal{L}^c_r(\vartheta_{ir})}{\partial \theta_{ir}} = - \frac{\partial \log f(y^*_{ir} - \theta_{ir})}{\partial \theta_{ir}}  = 1 - 2p_{ir}.
\] 
A closed form expression for the expectation of $p_{ir}$ is 
\citep{derooijbreemerwoestenburgbusing2022, jiao2016high}
\begin{eqnarray*}
\mathbb{E}(p| y, \theta, \bm{t})  = \left\{ \begin{array}{ll} 
\left[\frac{\exp(2t_{y} - 2\theta)} {2[\exp(t_{y} - \theta) + 1]^2}\right] / F(t_{y} - \theta)		& \mathrm{if}\ y = 1\\[4pt]
\left[\frac{2\exp(t_{(y-1)} - \theta) + 1} {2[\exp(t_{(y-1)} - \theta) + 1]^2} - \frac{2\exp(t_{y} - \theta)} {2[\exp(t_{y} - \theta) + 1]^2}\right]	/ \left(F(t_{y} - \theta)	- F(t_{(y-1)} - \theta)\right)	& \mathrm{if}\ 2\leq y < C \\[4pt]
\left[\frac{2\exp(t_{(y-1)} - \theta) + 1} {2[\exp(t_{(y-1)} - \theta) + 1]^2}\right] / \left( 1 - F(t_{(y-1)} - \theta) \right)			& \mathrm{if}\ y = C
\end{array}\right . 
\end{eqnarray*} 
where we used $y$ and $\theta$ instead of $y_{ir}$ and $\theta_{ir}$ for readibility. The expectation has to be evaluated at the current parameter estimates $\bm{\theta}$ and $\bm{t}$. Let us denote by $\xi_{ir}$ the expected value of the first derivative, that is
\[
\xi_{ir} = 1 - 2 \mathbb{E}(p_{ir}| y_{ir}, \theta_{ir}, \bm{t}_r).
\]
An upper bound for the (expectation of the) second derivative is given by any $\kappa \geq 1/4$. The $\xi$ and $\kappa$ can be used in the majorization function. This majorization function was derived by \cite{derooijbreemerwoestenburgbusing2022}. 

\paragraph{Combining majorization functions}

Our negative log-likelihood function is 
\[
\mathcal{L}(\bm{\theta}) = \sum_{r} \mathcal{L}_r(\bm{\theta})  = \sum_{i=1}^N \sum_{r=1}^R \mathcal{L}_{ir}(\theta_{ir}) . 
\]
We derived majorization functions for $\mathcal{L}_{ir}(\theta_{ir})$ for numeric, binary, and ordinal response variables, each having a least squares form. Because majorization is closed under summation, we have
\[
\mathcal{L}(\bm{\theta}) \leq \mathcal{M}(\bm{\theta}, \bm{\vartheta}),
\]
where
\[
\mathcal{M}(\bm{\theta}| \bm{\vartheta}) = \sum_i \sum_r \mathcal{M}(\theta_{ir}| \vartheta_{ir})  = \| \bm{Z} - \bm{1m}' - \bm{\Phi}\bm{B}\bm{V}'\|^2,
\]
is a least squares function. The matrix $\bm{Z}$ has elements $z_{ir} = \vartheta_{ir} - \frac{1}{\kappa^*}\xi_{ir}$, where $\kappa^* = \max(\frac{1}{4}, \sigma^{-2})$. The vector $\bm{m}$ contains the $m_r$ for $r \in \{\mathcal{N, D}\}$ and zeros for $r \in \mathcal{O}$. 

The choice of $\kappa^* = \max(\frac{1}{4}, \sigma^{-2})$ works fine except when the variance $\sigma^{2}$ is close to zero. As we do not know the value of this variance, we estimate it from the data in each iteration of the algorithm (see Section 3.1.4) and plug the estimate in the algorithm. An estimate close to zero happens, when the numeric response variable are approximated very well. In that case, $\kappa^*$ becomes a very large value and the working responses stay very close to the expected value of the previous iteration. In that case, the algorithm gets stuck. In our software, we implemented a warning when the estimated variance becomes very small, that is $\hat{\sigma}^{2} < 0.05$.

\subsubsection{Update of the Regression Weights}

To update the regression weights $\bm{B}$, we first define the auxiliary matrix $\tilde{\bm{Z}} = \bm{Z} - \bm{1m}'$. The least squares loss function can be written as 
\[
\|\tilde{\bm{Z}} - \bm{\Phi}\bm{BV}'\|^2 = 
\| \mathrm{Vec}(\tilde{\bm{Z}}) - \left(\bm{V} \otimes \bm{\Phi}\right) \mathrm{Vec}(\bm{B})\|^2 =
\| \tilde{\bm{z}} - \bm{H}\bm{b}\|^2, 
\]
where $\tilde{\bm{z}} = \mathrm{Vec}(\tilde{\bm{Z}})$, $\bm{b} = \mathrm{Vec}(\bm{B})$, and $\bm{H} = \bm{V} \otimes \bm{\Phi}$. This is a standard regression problem such that
\[
\bm{b}^+ = (\bm{H}'\bm{H})^{-1} \bm{H}\tilde{\bm{z}}.
\]
The computational burden can be reduced by noting that $\bm{H}'\bm{H} = \bm{I}_S \otimes \bm{\Phi}'\bm{\Phi}$, simplifying the computation of the inverse.

\subsubsection{Update of the Loadings}

To update the loadings, we use the same auxiliary matrix $\tilde{\bm{Z}}$ as above and need to minimize
\[
\|\tilde{\bm{Z}} - \bm{\Phi}\bm{BV}'\|^2, 
\]
under the restriction $\bm{V}'\bm{V} = \bm{I}$. This function can easily be minimized using the lower bounds described in \cite{tenberge1993}. Therefore, define the singular value decomposition
\[
\bm{B}'\bm{\Phi}'\tilde{\bm{Z}} = \bm{P}\bm{\Delta}\bm{Q}',
\]
such that an update for $\bm{V}$ is given by
\[
\bm{V}^+ = \bm{Q}_s\bm{P}'_S,
\]
where $\bm{Q}_S$ denote the singular vectors corresponding to the $S$ largest singular values and similarly for $\bm{P}_S$.
    
\subsubsection{Update of the intercepts}

To update the intercepts for numeric and binary response variables, we define the auxiliary matrix $\tilde{\bm{Z}} = \bm{Z} - \bm{\Phi}\bm{BV}'$, from which we only retain the columns for numeric and binary responses. With this auxiliary matrix, we need to minimize
\[
\|\tilde{\bm{Z}} - \bm{1m}'\|^2. 
\]
The solution is
\[
\bm{m}^+ = \tilde{\bm{Z}}'\bm{1}(\bm{1}'\bm{1})^{-1}.
\]

\subsubsection{Update of Residual Variance}

For updating $\sigma^2$, we only need the numeric responses. We therefore focus only on the columns related to $r \in \mathcal{N}$. For these columns, we compute the residuals
\[
\bm{E} = \tilde{\bm{Z}} - \bm{1m}' - \bm{\Phi}\bm{BV}',
\] 
for which we compute the variance to obtain an update of $\sigma^2$ as
\[
\frac{1}{N \cdot R_{\mathcal{N}} - 1}
\sum_{i = 1}^N \sum_{r \in \mathcal{N}} e^2_{ir} ,
\]
where $R_{\mathcal{N}}$ is the number of numeric response variables. 

\subsubsection{Update of the Quantifications}

For categorical predictor variables, we optimally scale the levels. For numeric variables, the quantified variable is simply the standardized predictor, as discussed in Section \ref{sec:model}, which remains constant throughout iterations. The relevant part of the majorization function for the transformations is
\[
\| \bm{Z} - \bm{1m}' - \bm{\Phi}\bm{A} \|^2 = \| \bm{Z} -  \bm{1m}' - \bm{\phi}_p\bm{a}'_p  - \bm{\Phi}_{(-p)}\bm{A}_{(-p)}\|^2,
\]
where $\bm{A} = \bm{BV}'$,  $\bm{\phi}_p$ is the $p$-th column of $\bm{\Phi}$, $\bm{a}_p$ is the column vector with the elements of the $p$-th row of $\bm{A}$, $\bm{\Phi}_{(-p)}$ is the matrix $\bm{\Phi}$ without the $p$-th column, $\bm{A}_{(-p)}$ is the matrix $\bm{A}$ without the $p$-th row. 

To find the transformation $\bm{\phi}_p = \varphi_p(\bm{x}_p)$, we define the indicator matrix $\bm{G}_p$ of size $N \times C_p$ and the vector of quantifications $\bm{w}_p$ of length $C_p$, so that  we can write the optimally quantified variable as
\[
\bm{\phi}_p = \varphi_p(\bm{x}_p) = \bm{G}_p \bm{w}_p,
\]
with - depending on the scaling level - constraints on $\bm{w}_p$. Let us first define the auxiliary matrix $\tilde{\bm{Z}} = \bm{Z} - \bm{1m}' - \bm{\Phi}_{(-p)}\bm{A}_{(-p)}$ then the majorization function becomes
\[ 
\| \mathrm{Vec}(\tilde{\bm{Z}}) - \left(\bm{a}_p \otimes \bm{G}_p \right) \bm{w}_p \|^2,
\]
a simple regression problem. Defining, $\bm{Q} = \bm{a}_p \otimes \bm{G}_p$ the unconstrained update is
\[
\bm{w}_p^+ =  (\bm{Q}'\bm{Q})^{-1} \bm{Q}'\tilde{\bm{z}}.
\]
For ordinal predictor variables, we need an extra step. In this extra step this update is projected onto the cone of admissible quantifications \citep{meulman2019ros}, i.e., the quantifications should have the correct order either increasing or decreasing.  This amounts to a weighted monotone regression \citep{deleeuw2005monotonic, busing2022monotone} of the $\bm{w}_p^+$ with weights equal to the observed frequencies of each of the response categories. Because the relationship can either be monotone increasing or monotone decreasing, we perform two of such monotone regressions and select the one that fits best. Note that these monotone regression problems are very small, as they only involve the number of categories of the ordered predictor variable.

In the last step of the optimal scaling process, we rescale $\bm{w}_p$ such that the mean of $phi_p$ is equal to zero and its variance equal to one. This is important, as we multiply the optimally scaled variable with a regression weight and without standardizing we would not be able to obtain unique estimates.

\subsection{Estimation of Thresholds}

The thresholds of the ordinal response variables  are not part of the canonical parameters. Where above we focused on minimizing $\mathcal{L}(\bm{\theta})$, that is,  
$\mathcal{L}(\bm{\theta}, \bm{t})$ with $\bm{t}$ fixed, now we will focus on minimizing $\mathcal{L}(\bm t)$, that is,  
$\mathcal{L}(\bm{\theta}, \bm{t})$ with $\bm{\theta}$ fixed. 
Because of the local independence assumption, we can estimate the thresholds separately for each response variable. For estimation of the thresholds for response variable $r \in \mathcal{O}$, we use standard maximum likelihood estimation where we fixed the other estimates.  

\subsection{Summary}

To summarize our algorithm. The algorithm needs starting values, that we derive using a reduced rank regression model where we take all predictors and responses as numeric variables. Starting with these estimates, we take the following steps:
\begin{itemize}
\item Compute the new working response variables;
\item Update the quantifications ($\phi_p$);
\item Update the weights ($\bm{B})$
\item Update the loadings ($\bm{V}$)
\item Update the intercepts for numeric and binary responses ($m$);
\item Update the threshold ($\bm{t}$). 
\end{itemize}
These steps are repeated until the decrease of the negative log-likelihood is smaller than a pre-set convergence criterion (e.g., $1 \times 10^{-6}$). The algorithm results in a unique solution. The order of the steps described above can be changed, that would lead to the same estimates.

\section{Simulation Study}\label{sec:simstudy}

To evaluate our algorithm, we conducted a simulation study. As discussed in the introduction, \cite{luo2018leveraging} proposed a reduced rank model for mixed numeric, binary, and count variables response variables and numeric predictors. We propose a reduced rank model for mixed numeric, binary and ordinal response variables and mixed predictor variables. We can therefore compare the algorithms for a mixture of binary and numeric response variables with numeric predictors.  

We followed the set-up of \cite{luo2018leveraging}. We consider their Model 1, a low-dimensional example, with a few adaptions. We set $N = 500$, $P = 8$, $R = 8$, and $S = 2$. Among the  responses,  4 of them are generated from Normal distribution and 4 from the Bernoulli distribution. The predictor matrix is constructed by generating its entries as independent and identically distributed random samples from the standard normal distribution $\mathcal{N}(0,1)$. The coefficient matrix $\bm{B}$ is an orthogonal matrix from the QR decomposition of a random  matrix filled with $\mathcal{N}(0,1)$ entries, and all entries in $\bm{V}$  are samples from the uniform distribution $\mathcal{U}(-1,1)$. We set the intercept vector equal to $\bm{0}$. The numeric responses are drawn from the normal distribution with mean equal to the canonical term and variance equal to one. The binary responses are drawn from the Bernoulli distribution. 

To evaluate the performance we compare the population $\bm{BV}'$ with the estimated $\hat{\bm{B}}\hat{\bm{V}}'$ using the root mean squared error metric (Rmse). The number of replications equals 250. On each generated data set, we fitted the reduced rank model with the algorithm of \cite{luo2018leveraging} and our algorithm. For each algorithm, we thus obtain 250 Rmse measures. We can compare the recovery of both algorithms using boxplots. 

Our methodology allows for categorical predictors and ordinal responses. To test our algorithm under these circumstances, we added three conditions. First, considering ordinal predictor variables, we discretized the numeric predictors into five categories based on the .2, .4, .6, and .8 quantiles. For generating the data, we used the average value of the numeric predictors within each of these categories, whereas for data analysis we coded the predictors using the integers 1 to 5. 

Second, considering ordinal responses, we generated two different types of data. In the first scenario, 4 binary and 4 ordinal responses were drawn, whereas in the second scenario four numeric and four ordinal variables were drawn. The ordinal response variables have four categories, and were generated by drawing from a multinomial distribution. The estimated thresholds used for deriving the probabilities were set to -1, 0, and 1. 

Also for these three extra conditions, we compare the population $\bm{BV}'$ with the corresponding estimates using the root mean squared error.  We depict the Rmse's from the 250 replications using a boxplot. 

\begin{figure}
    \centering
    \includegraphics[width = 0.8\textwidth]{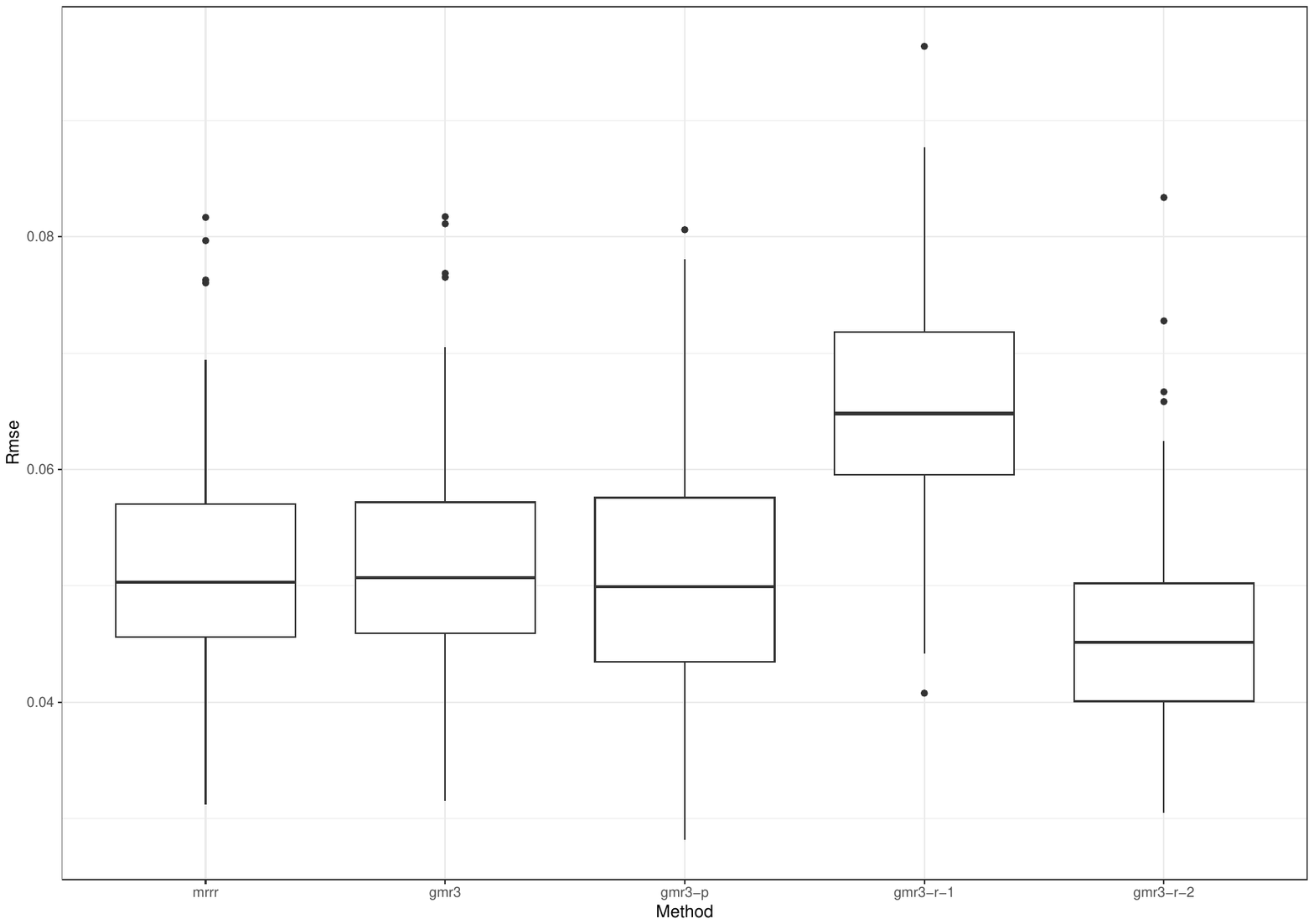}
    \caption{Results of simulation study. Different methods are compared in terms of Root mean squared error (Rmse). The first method is the mixed reduced rank regression algorithm of Luo and colleagues, the second is our algorithm. These two are applied to data with numeric predictors and a mix of numeric and binary responses. In the third (gmr3-p), we changed the predictors to be ordinal variables, so that the optimal scaling is employed. In the fourth and fifth method, we generated ordinal responses: in gmr3-r1 the responses are a mixture of binary and ordinal responses, whereas in gmr3-r2 they are a mixture of numeric and ordinal responses.}
    \label{fig:simresults}
\end{figure}

The results are shown in Figure \ref{fig:simresults}. We see that our algorithm performs the same as the algorithm of \cite{luo2018leveraging}. Furthermore, we can conclude that changing the predictors into ordinal variables with the use of optimal scaling in the algorithm does not change the performance. The latter two boxplots, show the results with ordinal response variables. When the response variables are a mixture of binary and ordinal variables, the recovery is a bit worse, but still good, whereas the recovery is a little better -- compared to a mixture of numeric and binary -- when the responses constitute a mixture of numeric and ordinal variables. 

\section{Model Selection for GMR$^3$}\label{sec:modsel}

Before we show an application on empirical data, we like to set out our thoughts on model selection for our generalized mixed reduced rank regression model. We discuss several existing methods and how they can be used in our framework.  

\subsection{The overall number of parameters}

The proposed class of models can be very complex in terms of number of parameters to estimate. 
This number depends on the number $R$ of the responses, the  number $P$ of the predictors, the reduced-rank $S$ as also the number of categories of discrete predictors, and the number of categories of the ordinal responses. 

As described in section \ref{sec:model}, the $R$ response variables are partitioned into the set $\mathcal{N}$ of numeric variables, the set $\mathcal{B}$ of binary variables, and the set $\mathcal{O}$ of ordinal variables. The $P$ predictors ($p = 1\ldots, P$) are partitioned in two sets:  a set of numeric variables $\mathcal{N}_p$ and a set of discrete variables $\mathcal{D}_p$.

For discrete predictors, the number of parameters estimated in the optimal scaling is equal to the number of categories \citep{deleeuw2005monotonic}. However, we restrict the quantifications to have mean zero and variance one, and therefore the number of parameters involved is $C_p - 2$. For ordinal responses, we estimate $C_r - 1$ thresholds, whereas for numeric and binary responses we estimate 1 intercept. The overall number of parameters to estimate can be obtained in the following way:

\begin{equation}
    {\cal K} = (P + R - S) \times S + \left[\sum_{p \in {\cal D}_P} C_p - 2 \right]  + \sum_{r \in \{{\cal N},{\cal D}\}} 1 + \left[\sum_{r \in {\cal O}} C_r - 1 \right]
\end{equation}

\noindent
where $C_p$ is the number of categories of the $p$-th discrete predictor and $C_r$ is the number of categories of the $r$-th ordinal response variable.

\subsection{Model selection procedure}\label{sec:modelselection}

Model selection requires the selection of both the predictors and the optimal dimensionality \footnote{The number of parameters to estimate depends also on the number of response variables but we will not select the response variables, all are included.}.

We adopt a two-step model selection procedure:
\begin{itemize}
    \item[I.] We fix the set of predictors and then we select the optimal dimensionality.
    \item[II.] Once the dimensionality is fixed, we verify which predictors are significant.
\end{itemize}

For step I, we identify a set of competing reduced-rank models with an increasing level of complexity. Section (\ref{complexity}) describes how to select the model complexity such to minimize the expected prediction error when using the model for fresh data with unknown response measurements. 

For step II, once the dimensionality is fixed in step I, we compute the $(1-\alpha)\%$ bootstrap confidence regions of the parameter estimates: the significant predictors are those for which the region does not include the origin (i.e., vector with zeros).

\subsection{The model complexity}
\label{complexity}

Consider $K$ competing models of the set ${\cal M} = \{{\cal M}_0, {\cal M}_1, \dots, {\cal M}_k, \dots, {\cal M}_K\}$, all generated by restricting the overall parameter vector space in decreasing order into the vector  ${\bm{\theta}}_k$. The number of parameters for model ${\cal M}_k$ is given by ${\cal K}_k$. 
The model ${\cal M}_0$ is the null model which includes only the intercept and threshold parameters, thus there is no contribution of the predictors. The number ${\cal K}_k$ can be also understood as the {\it tuning parameter} that governs the trade-off between bias and variance\footnote{The variance refers to the amount by which the fitted model would change if it is estimated using a different training sample, whereas the bias refers to the error due to a much simpler mode. Complex models lead to overfitting and cannot be generalized to fresh data (propagation error). A simple model yields to model inadequacy error. The goal is to minimize the expected prediction error which simultaneously achieves low variance and low bias.} \citep{vapnik1995, vapnik1998}. 

The aim is to find an optimal value $k^*$ which yields the smallest prediction error of fresh data with unknown response measurements. Such prediction error can be estimated {\it indirectly} by making an adjustment to the training error account for the bias due to overfitting. Another approach is to {\it directly} estimate the test error using cross-validation (\cite{hastie2009elements}).

\subsection{Goodness of Fit-based Model Selection Criteria}

Akaike Information Criterion (AIC) and Bayesian Information Criterion (BIC) are commonly used for model selection \citep{hastie2009elements}. Another choice is given by Mc Fadden R-squared goodness of fit measure \citep{mcfadden1974}. 

The general idea is to penalize the goodness of fit measure of the model, fitted using a training sample, by the number of the estimated parameters. It is a matter of evaluating if it is worth to add more parameters with respect to the improvement of goodness of fit a more complex model provides. 

McFadden’s adjusted R-squared with respect to the standard R-squared measure takes into account the complexity of the model in the following way: 

\begin{equation}
\label{adjR2}
R_a^2({\cal M}_k) = 1 - \frac{\hat{\mathcal{L}}(\bm{\theta}_k, \bm{t}) + {\cal K}_k}{\hat{\mathcal{L}}(\bm{\theta}_0, \bm{t})},
\end{equation}

\noindent
where $\hat{\mathcal{L}}(\bm{\theta}_k, \bm{t})$ is the achieved minimum of the negative log likelihood for the model ${\cal M}_k$ and similarly $\hat{\mathcal{L}}(\bm{\theta}_0, \bm{t})$ is that quantity for the null model ${\cal M}_0$. Adjusted McFadden $R_a^2$ penalizes the goodness of fit of the current model by the number of parameters to be added with respect the null model\footnote{Note that negative McFadden’s adjusted R-squared is possible.}. The optimal model choice maximizes the $R_a^2({\cal M}_k)$ over all $k$ in the set ${\cal M}$. 

AIC is based on the entropic or information-theoretic interpretation of the maximum likelihood method as well as the minimization of the Kullback-Leibler (K-L) information quantity. AIC for any model ${\cal M}_k$ can be defined as:

\begin{equation}
\label{aic}
    AIC({\cal M}_k) =  2 \hat{\mathcal{L}}(\bm{\theta}_k, \bm{t}) + 2{\cal K}_k.
\end{equation}

\noindent
The first term $2\hat{\mathcal{L}}(\bm{\theta}_k, \bm{t})$ in AIC is twice the negative log likelihood and it acts as a measure of lack of fit to the data, consequently smaller values will be preferred. The second term $2{\cal K}_k$ acts as a penalty term which penalizes models having many parameters. The aim is to reach a balance between the lack of fit and the model complexity: models with smaller AIC values  indicate a better balance. The optimal model choice minimizes the $AIC({\cal M}_k)$ over all $k$ in the set ${\cal M}$. 

BIC is an alternative to AIC and is based on an asymptotic Bayesian argument\footnote{The idea behind BIC is that the maximum likelihood estimators for arbitrary nowhere vanishing a priori distributions can be obtained as large-sample limits of the Bayes estimators. Thus a suitable modification of maximum likelihood is searched, through the analysis of the asymptotic behaviour of Bayes estimators under a special, not absolutely continuous, class of priors.}. Having a finite number of models in the set ${\cal M}$, selecting the one with the highest marginal log-likelihood in large samples, is equivalent to minimizing the following measure for all $k$ in the set ${\cal M}$:

\begin{equation}
\label{bic}
    BIC({\cal M}_k) = 2 \hat{\mathcal{L}}(\bm{\theta}_k, \bm{t}) + \log(N){\cal K}_k,
\end{equation}

\noindent
where $N$ refers to the sample size. The model which minimizes BIC corresponds to the model with the highest posterior probability. Due to the larger penalty of $\log(N)$ on the model complexity as opposed to $2$ for AIC, BIC often selects a sparser model compared to AIC.

\subsection{Model Selection by Cross-Validation}

Model selection can be performed by a direct estimate of the prediction error through $L$ times repeated $V$-Fold Cross-validation. The sample is repeatedly partitioned into $V$ folds, each at turn is used to evaluate the model fitted using the remaining $(V-1)$ ones. 
We derive the $V$-Fold Cross-validation estimate of the prediction error for each model ${\cal M}_k$ in the set ${\cal M}$ in the following way:

\begin{equation}
\label{cv}
    CV({\cal M}_k) = \frac{1}{L}\sum_{l = 1}^L \frac{1}{V} \sum_{v=1}^V \frac{1}{n_v} \sum_{i=1}^{n_v} \sum_{r = 1}^R {\cal L}_v({\bm{\hat \theta}_k)}_{ir},
\end{equation}

\noindent
where 
${\cal L}_v({\bm{\hat \theta}_k)}_{ir}$ is the loss function for the $i$-th individual and the $r$-th response variable evaluated in the $v$-fold for the model ${\cal M}_k$ which parameters are estimated using the remaining $(v-1)$ folds. Typical choices are $V=5$ or $V=10$. The optimal model choice is the model which minimizes the $CV({\cal M}_k)$ over all $k$ in the set ${\cal M}$ or the one that fits within one-standard error of the minimum prediction error estimate.

\subsection{Bootstrap}

We use the bootstrap \citep{efron1979bootstrap, efron1986bootstrap, davison1997bootstrap} to obtain confidence regions for the regression weights and loadings. For regression models, researchers can choose between randomly drawing pairs, that is both the explanatory and response variables, or drawing residuals. The latter assumes that the functional form of regression model is correct, that the errors are identically distributed and that the predictors are fixed \citep{davison1997bootstrap}. For our generalized mixed reduced rank regression method, we draw pairs of sets of explanatory and response variables, to avoid the dependency upon these assumptions. This sampling scheme also takes into account that there are dependencies among the response variables of a participant and the bootstrap automatically adapts to these dependencies. This resembles the so-called clustered bootstrap for nested or hierarchical data \citep{sherman1997comparison, deen2020clusterbootstrap}. 

The balanced bootstrap can be used to ensure that every participant appears exactly ``the number of bootstrap'' times in the bootstrap samples, in contrast to randomly drawing bootstrap samples from the parent sample. \cite{davison1997bootstrap} show that the balanced bootstrap results in an efficiency gain. 

Bootstrap confidence ellipses can be visualized by data ellipses, as discussed by \cite{friendly2013elliptical}. To verify whether a predictor variable has a significant contribution, we verify whether the ellipse of predictor $p$ includes the origin. Similarly, to verify whether a response is predicted well from the set of predictors, we verify whether the ellipse of the loadings for response $r$ includes the origin of the space.

\section{Application}\label{sec:applications}

\subsection{Eurobarometer Surveys}\label{sec:data}

We will use data from the European Commission’s Eurobarometer Surveys of January-February 2023 \citep{ZA7953}, with a specific focus on residents of the Netherlands, to illustrate our methods.
For $N = 837$ Dutch inhabitants we have their opinion on various issues related to Europe, including unification, institutions and policies. We will consider response variables of ordinal and binary type as well as predictors of binary, ordinal and numerical type. For discrete predictor variables, we use optimal scaling to quantify the categories. Hereby the list of variables with their scale of measurement and original coding:

\begin{itemize}
\item Ordinal response variables:
\begin{enumerate}
\item[CI:] The interests of the Netherlands are taken into account in the European Union. Ordinal scaled categories: 'Strongly Disagree' (SD = 1), 'Disagree' (D = 2), 'Agree' (A = 3), 'Strongly Agree' (SA = 4).
\item[MW:] Every EU member state should have a minimum wage for workers. Ordinal scaled categories: 'Strongly Disagree' (SD = 1), 'Disagree' (D = 2), 'Agree' (A = 3), 'Strongly Agree' (SA = 4).
\item[FS:] The EU has taken a series of measures in response to Russia's invasion of Ukraine. To what extent do you agree or disagree with providing financial support to Ukraine? Ordinal scaled categories: 'Strongly Disagree' (SD = 1), 'Disagree' (D = 2), 'Agree' (A = 3), 'Strongly Agree' (SA = 4).
\item[DI:] More money needs to be spent on defense in the EU. Ordinal scaled categories: 'Strongly Disagree' (SD = 1), 'Disagree' (D = 2), 'Agree' (A = 3), 'Strongly Agree' (SA = 4).
\item[RE:] Reducing oil and gas imports and investing in renewable energy is important for our overall security. Ordinal scaled categories: 'Strongly Disagree' (SD = 1), 'Disagree' (D = 2), 'Agree' (A = 3), 'Strongly Agree' (SA = 4).
\end{enumerate}
\item Binary response variables:
\begin{enumerate}
\item[T:] Do you rather or do you not have confidence in the European Union? Rather trust = 1, Rather not trust = 0. 
\item[FE:] What is your opinion about further expansion of the EU to include other countries in the future? Pro = 1, Against = 0. 
\end{enumerate}
\end{itemize}

\begin{itemize}
\item Binary predictor
\begin{enumerate}
\item[G:] Gender (Male = 0, Female = 1). 
\end{enumerate}
\item Ordinal predictors:
\begin{enumerate}
\item[PA:] Political Alignment with three ordinal categories 'Left' = 1, 'Center' = 2, 'Right' = 3.
\item[U:] Urbanization with three ordinal categories 'Rural Area/Village' = 1, 'Small/Middle Town' = 2, 'Large Town' = 3).
\item[E:] The highest level of education attained with ordinal categories 'Pre-primary' = 1, 'Primary' = 2, 'Low Secondary School' = 3, 'Up Secondary School' = 4, 'Post Secondary School' = 5, 'Tertiary' = 6, 'Bachelor' = 7, 'Master' = 8, and 'Doctorate' = 9.
\end{enumerate}
\item Numeric predictor
\begin{enumerate}
\item[A:] Age (from 15 to 75 years), standardized to have mean 0 and variance 1.
\end{enumerate}
\end{itemize}

\subsection{Model Selection}

\subsubsection{Goodness of Fit}

\begin{table}[t]
\centering
\begin{tabular}{rrrrrrr}
  \hline
   $S$ & $\mathcal{L}$ & ${\cal K}$ & AIC & BIC & $R_a^2$ \\ 
  \hline
  1 & 5055.49 & 37 & 10184.98 & 10359.98 &  0.033 \\ 
  2 & 5006.52 & 46 & 10105.03 & 10322.60 &  0.041 \\ 
  3 & 4990.00 & 53 & 10086.01 & 10336.69 &  0.043 \\ 
  4 & 4988.96 & 58 & 10093.91 & 10368.24 &  0.042 \\ 
  5 & 4988.25 & 61 & 10098.51 & 10387.03 &  0.042 \\ 
   \hline
\end{tabular}
\caption{Model fit statistics for the Eurobarometer data for the set of models including all predictors for different dimensionalities ($S$).}
\label{tab:dimselection}
\end{table}

Model selection criteria can be based on goodness of fit measures. Both Akaike and Bayesian Information criteria AIC and BIC include the penalization factor accounted for the model complexity as measured by the number of parameters to estimate. For this purpose, Mc Fadden R-squared measure $R^2$ has also the adjusted version $R_a^2$. All these measures are evaluated for all models with an increasing dimensionality $S$.  With larger dimensionality, the number ${\cal K}$ of parameters to estimate also increases. Model selection results are shown in Table \ref{tab:dimselection}. In our application, the BIC suggests a two-dimensional (i.e., $S=2$) model with ${\cal K}=46$ parameters whereas, as often is the case, the AIC proposes a less parsimonious three dimensional model with ${\cal K}=53$ parameters. The adjusted  Mc Fadden R-squared measure also suggests three dimensional model, although the differences between the 2, 3, 4, and 5 dimensional models are small. 

\subsubsection{Cross validation}

\begin{figure}
    \centering
    \includegraphics[width = 0.8\textwidth]{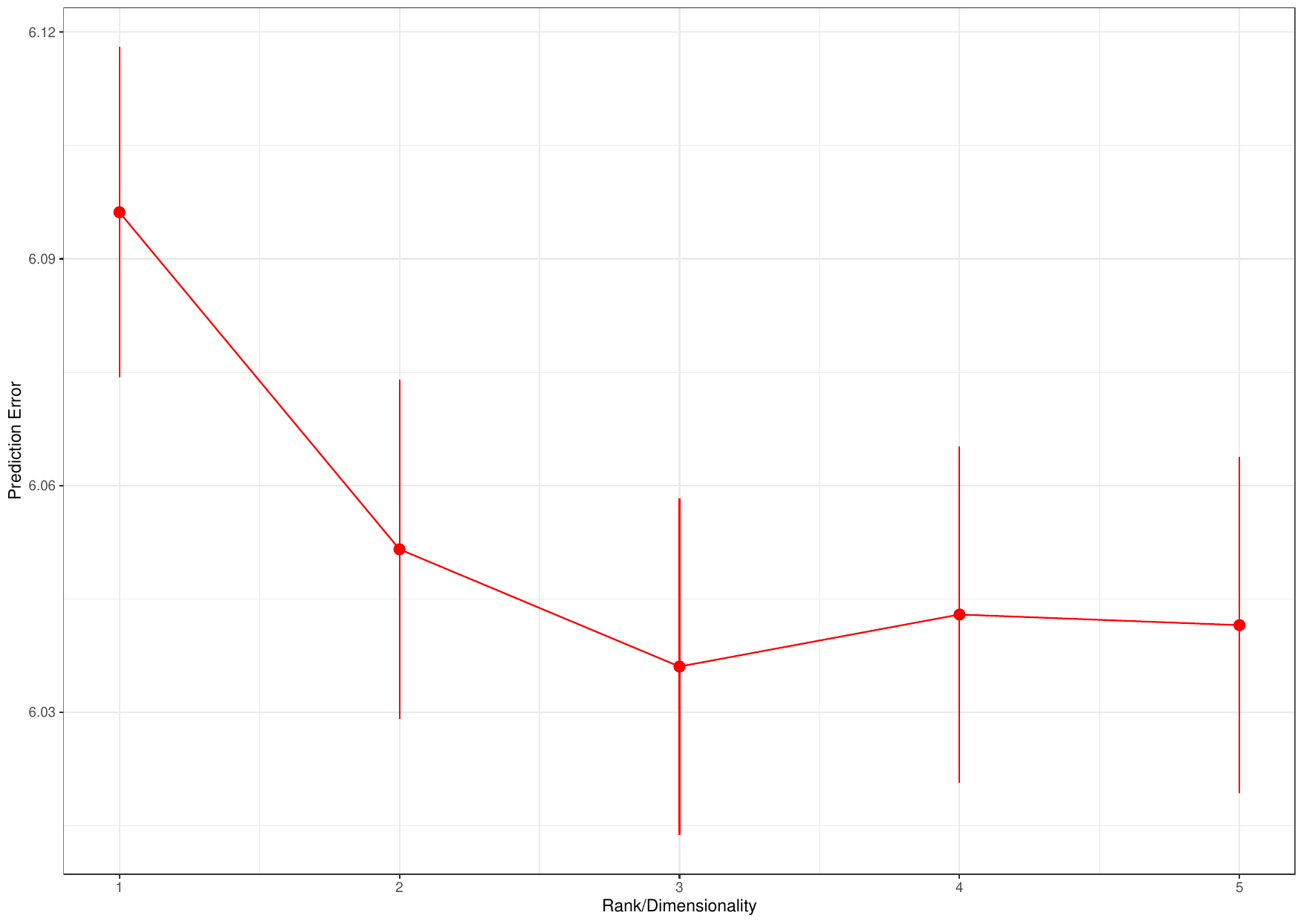}
    \caption{Cross validated prediction error for dimensionalities from one till five. Average prediction error (points) plus and minus one standard error (bars) are shown.}
    \label{fig:xval}
\end{figure}

We did ten times repeated $10$-fold cross validation to select the optimal dimensionality (or rank). The results are visualized in Figure \ref{fig:xval}. The plot shows the average prediction error per participant in the validation sets, against the rank of the model. Also included are error bars, that represent the standard error. It can be verified that the rank three model has the lowest prediction error. However, the rank two model (i.e., a more parsimonious model) falls within the one-standard error range. 

From the goodness of fit analysis and the cross validation results, we conclude that the two-dimensional model provides a good representation of the data.

\subsubsection{Bootstrap}

\begin{figure}
    \centering
    \includegraphics[width = .9\textwidth]{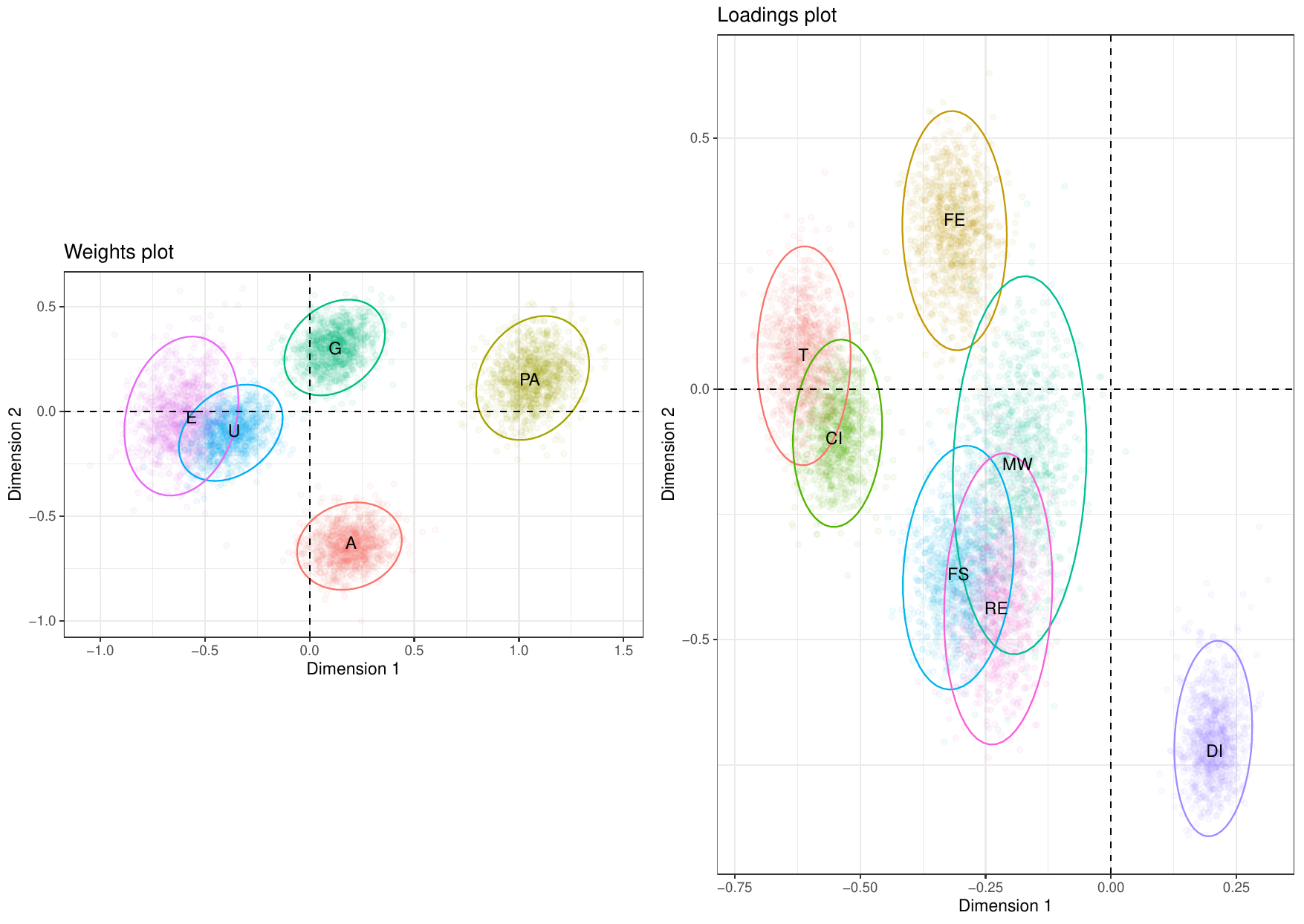}
    \caption{Results of 1000 bootstraps. On the left, the results for the regression weights ($\bm{B}$), and on the right the results of the loadings ($\bm{V}$). For abbreviations of the variables see Section \ref{sec:data}.}
    \label{fig:bootres}
\end{figure}

To investigate the contributions of the predictor variables, we performed a bootstrap analysis. One thousand bootstrap samples were drawn and for each sample the model was fitted. In Figure \ref{fig:bootres}, we show the results of the bootstrap. We separate the information in two graphs, one for the weights ($\bm{B}$) and one for the loadings ($\bm{V}$). 

For the regression weights we can conclude that all predictor variables have a significant contribution in the model, that is, non of the 95\% confidence ellipses include the origin. The same conclusion can be drawn for the loadings. 

\subsection{Interpretation}

\begin{figure}
    \centering
    \includegraphics[width = .9\textwidth]{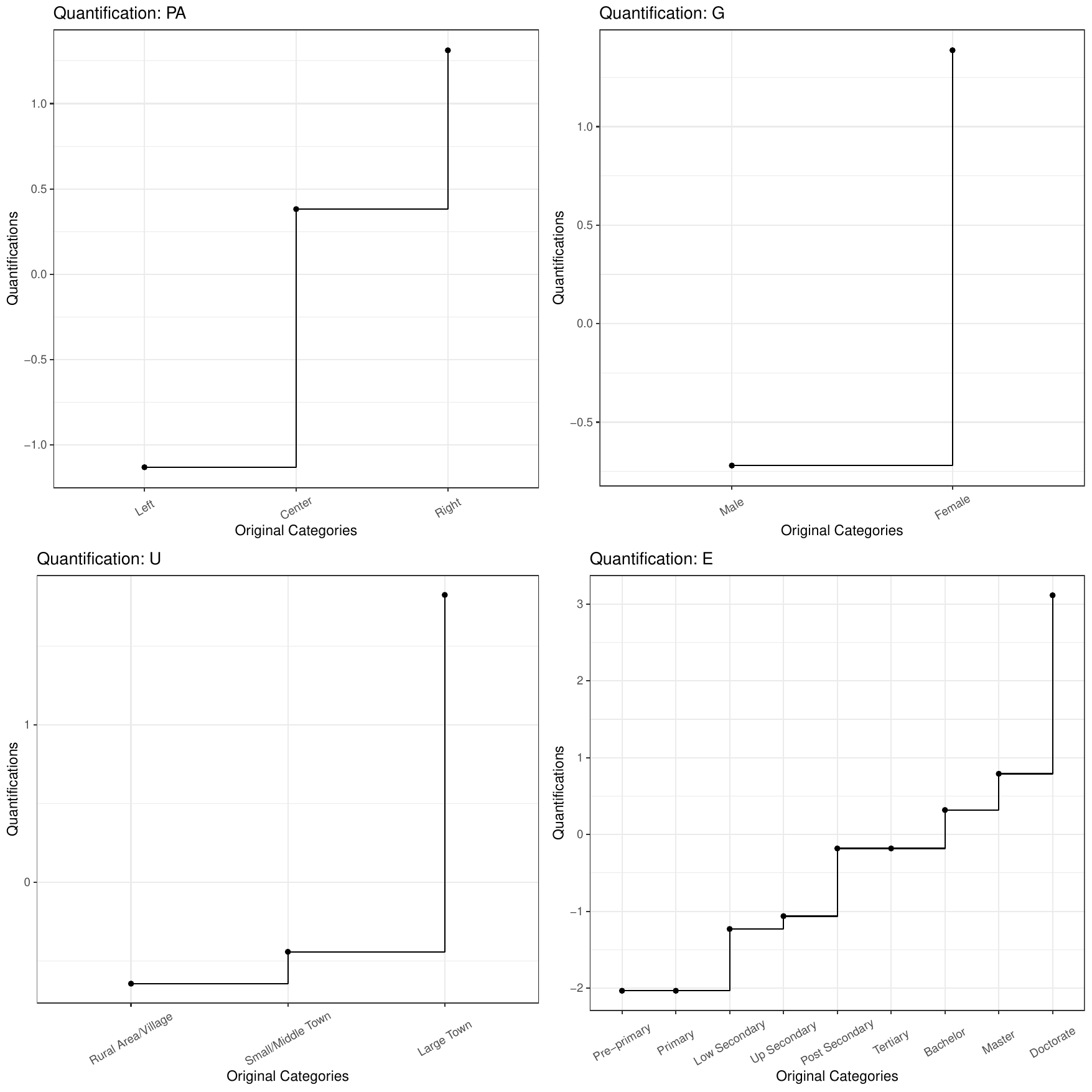}
    \caption{Data transformations of predictors. Nominal or ordinal quantifications on the vertical axis versus original categorical values on the horizontal axis.}
    \label{fig:mainfig}
\end{figure}

We start to show the optimally quantified variables in Figure \ref{fig:mainfig}. In this figure, it can be seen that for the gender the quantification for man is -0.72, while that for woman is 1.39. For political allignment, the quantifications are -1.13 for left, 0.36 for middle, and 1.31 for right. The difference between left and middle is approximately 1.5 and the difference between right and middle is approximately one unit. These differences can be important for detailed interpretation, as the coefficients are often interpreted as changes in the expected value or the estimated log odds with a unit change in the optimally scaled predictor. For urbanization, we see that there is not much difference for the rural area/village (-0.64) and the small/middle town (-0.44), but that the large town stands out (1.82). For education, we see that the first two levels have about equal quantifications, also the 3rd and 4th levels have similar quantifications. Then there is a cluster of post secondary, tertiary, bachelor and master, and finally there is a doctoral degree that stands out.

Figure \ref{fig:bootres}, also shows the parameter estimates of the weights and loadings. For example, the regression weights for age (A) are 0.65 and -0.13, which can be verified in the plot. Similarly, the loadings for trust (T) are -0.36 and -0.50. 

We can conclude from this Figure that the effects of the predictors urbanization (U) and education (E) are similar, although the effect for education is stronger (larger distance form origin). The effect of political allignment (PA), however, is opposite to that of urbanization (U) and education (E). 

For the response variable side of the model, we can see that the predictors influence the two response variables trust (T) and countries interest (CI) in the same way, because their loadings are very similar.
The same reasoning holds for the three response variables minimum wage (MW), financial support for  Ukraine (FS), and renewable energy (E). The response variables defense investments (DI) and future enlargement (FE) are a bit isolated, so predictors influence these responses in a different way. The effects on the latter two response variables are opposite in some sense. We see that the reduced rank model implies the responses to be associated. 

We can verify that all response variables are described by the predictors to some extent, that is, none of the confidence intervals of the responses includes the origin. Some response variables are closer to the origin (i.e., minimum wage (MW)) whereas others (i.e., countries interest (CI) and trust (T)) are further away from the origin. The distance to the origin represents the strength of discrimination for that response variable.

From the estimates of the regression weights and the loadings, we can compute the implied parameter estimates, $\hat{\bm{B}}\hat{\bm{V}'}$, that is

\begin{table}[H]
\centering
\begin{tabular}{rrrrrrrr}
  \hline
 & T & FE & CI & MW & FS & DI & RE \\ 
  \hline
  A & -0.16 & -0.27 & -0.05 & 0.06 & 0.17 & 0.50 & 0.23 \\ 
  PA & -0.63 & -0.28 & -0.60 & -0.22 & -0.38 & 0.11 & -0.31 \\ 
  Gr & -0.06 & 0.06 & -0.10 & -0.07 & -0.15 & -0.19 & -0.16 \\ 
  U & 0.21 & 0.08 & 0.21 & 0.08 & 0.14 & -0.01 & 0.12 \\ 
  E & 0.34 & 0.17 & 0.31 & 0.11 & 0.18 & -0.10 & 0.14 \\ 
   \hline
\end{tabular}
\end{table}
that shows in the rows the predictor variables and in the columns the response variables. The coefficients can be interpreted as in standard binary or ordinal logistic regression models.

Suppose, we have a 70 year old woman (quantification for age is 1.25 and for woman is 1.39), who is politically left oriented (quantification is -1.13), who lives in a rural area (quantification is -0.64) and has a bachelor degree (quantification is 0.32). From these optimally quantified categories of the predictor variables and the estimated coefficients, we can derive the expected value for the first response variable. Note this first response variable is binary, so we compute

\begin{eqnarray*}
\theta_{i1} &=& 0.45 + 1.25 \times (-0.16) -1.13 \times (-0.63) \\
&& + 1.39 \times (-0.06)  - 0.64 \times 0.21 + 0.32 \times 0.34 = 0.85,
\end{eqnarray*}
where 0.45 is the estimated intercept for the first variable (not shown before). With this value of the canonical parameter, we can compute the probability that this participant trusts the European Union as an institution, that is,
\[
\pi_{i1} = \frac{\exp(0.85)}{ 1 + \exp(0.85) } = 0.70
\]

Similarly, for variable 3, which is an ordinal variable, we compute
\begin{eqnarray*}
\theta_{i3} &=& 1.25 \times (-0.05) -1.13 \times (-0.60) \\
&& + 1.39 \times (-0.10)  - 0.64 \times 0.21 + 0.32 \times 0.31 = 0.45
\end{eqnarray*}

We have to compare this canonical term to the estimated thresholds for this response variable, that is $t_{1} = -2.57$, $t_2 = -0.91$, and $t_3 = 1.80$. As the canonical parameter falls within the threshold $t_2$ and $t_3$, we classify this person in class 3, i.e., she tends to agree with this question. Alternatively, we can compute the estimated probabilities that this person responds with each of the four answer categories. These probabilities are 0.05 for strongly disagree, 0.16 for disagree, 0.59 for agree, and  0.21 for strongly agree. Again, this person tends to agree. 

Some general conclusions about this Dutch participants that can be derived from the implied coefficients. As the participants grow older (Age), they have less trust in the European Union as an institution, they do not think the EU should be enlarged, believe that their country's interests are not sufficiently respected, tend to agree with a minimum EU wage, the financial support of Ukraine, more defense investments and investments in renewable energy. 

As Dutch people are more right wing (PA), they have less trust in the European Union as an institution, they do not think the EU should be enlarged, believe that their country's interests are not sufficiently respected, tend to disagree with a minimum EU wage, the financial support of Ukraine, investments in renewable energy, but tend to agree with more defense investments.

Higher educated participants have more trust in the European Union as an institution, they do think the EU should be enlarged, believe that their country's interests are sufficiently respected, tend to agree with a minimum EU wage, the financial support of Ukraine, investments in renewable energy, but tend to disagree with more defense investments.

\subsection{Comparison to standard regression models for each of the responses}

We also analyzed the data by running standard regression methods for each of the responses separately. For Trust in the European Union (T) and further expansion of it (FE), this amounts to fitting binary logistic models and for the others (CI. MW, FS, DI, and RE) proportional odds regression models. For the categorical predictor variables we used dummy coding as is usual, with the first category as baseline. We estimate the models and use a bootstrap procedure for finding the standard errors of the parameters. To make a fair comparison, or our GMR$^3$ model, we derived coefficients similar in meaning to those of the separate models. The results are described in detail in Appendix A. We describe here some general conclusions.

Although the total negative log-likelihood is lower than the one obtained with our methodology the sum of AIC and BIC statistics is higher than the AIC and BIC we obtained. This is mainly because of the number of parameters, which for the separate models equals 115, while for our model is 46. The estimated coefficients of the separate models do not take into account the ordered nature of the predictors (see Table \ref{tab:separatemodels} and compare to Table \ref{tab:impliedgmr3}). Some of the estimated coefficients become really large in the separate models, for example, the effects of the dummy variables for eduction on the response variables MW, FS, DI, and RE. Such large estimates usually point to very unstable results. The implied corresponding estimates of our model are much better. We can also witness the instability of the estimates in the bootstrap standard errors, shown in Table \ref{tab:bootseparate}. Some of these standard errors are really huge. As a comparison, we showed the standard errors of the corresponding estimates from our model in Table \ref{tab:bootimpliedgmr3}. By fitting one low rank model instead of separate regression model, we obtain better estimates that are much stabler. Furthermore, from our results in Figure \ref{fig:bootres} we directly see that the effects from the predictors on, for example, FS and RE are very similar where such a conclusion is very difficult to obtain from the separate fits.

\section{Conclusions and Discussion}\label{sec:disc}

In this paper, we developed a reduced rank regression model for mixed type of response variables and mixed type of predictor variables. We named our method GMR$^3$, the generalized mixed reduced rank regression model. Earlier work considered reduced rank regression models for a single type of response variables. For numeric responses, the reduced rank model was first described by \cite{anderson1951estimating} and further developed in the 70s and 80s of the previous century \citep{anderson1951estimating, izenman1975reduced, tso1981reduced, davies1982procedures, wollenberg1977redundancy}. Compared to a multivariate regression model, the reduced rank structure is imposed to keep the number of parameters relatively small. The reduced rank model was generalized to response variables of the exponential family \citep{yee2003reduced, yee2015}, to binary variables \citep{derooij2023new}, and to ordinal variables \citep{derooijbreemerwoestenburgbusing2022}. Few of these, however, consider mixed type of response variables. In this paper, we combined the original numeric model, with that for binary and ordinal response variables, to obtain a reduced rank model for a mixture of response variables. These types of variables often occur in the social sciences, like economics, psychology, education, political science, but also health related sciences.  

Furthermore, we also considered different types of predictor variables, whereas the earlier models could only deal with numeric variables. Therefore, discrete predictor variables have to be coded as dummy variables. We, however, used optimal scaling to quantify the discrete predictors. Whereas there is a lot of knowledge on optimal scaling within least squares problems \citep[see, for example, ][]{gifi1990nonlinear}, optimal scaling has not been applied within maximum likelihood estimation. We show an estimation procedure, where the quantifications are chosen such that they maximize the log likelihood of the final model. This was made possible by the employment of the MM algorithm we developed. 

We developed a majorization-minimization algorithm for maximum likelihood estimation of the parameters of the model. We showed that the negative log likelihood for the different response variables can be majorized by a least squares function. For minimizing least squares functions a lot of knowledge is available \citep[cf.][]{tenberge1993}. The algorithm monotonically converges to a minimum. Because each of the original functions per response variable is convex, the sum of these is also convex. Therefore, the attained minimum is also the global minimum. In our application, the algorithm turned out to be quite fast, that is, only a few iterations are needed for convergence. For our cross validation and bootstrap functions we used warm starts to speed up the procedures, where the starting values are equal to the final solutions obtained on the complete data set. We discussed in Section \ref{sec:mm} that the algorithm might have difficulties in finding the minimum when the estimated variance of the residuals for the numeric response variables is small. In our application and simulation studies we did not encounter this issue. Other properties of the algorithm still have to be investigated.  

In Section \ref{sec:modelselection}, we proposed a step-wise model selection procedure where first the rank or dimensionality is chosen and thereafter a bootstrap procedure is employed to verify which predictor variables have an effect on the responses. Such a step-wise procedure is efficient in terms of the number of models to be fitted. However, it does not guarantee to find the best model. Therefore, it would be better to fit models wit varying sets of predictors in all possible dimensionalities. Such a procedure requires fitting many models and can be time consuming. Still, if we would use the full grid of possible models, different statistics of model selection (AIC and BIC, for example) would probably point towards different best models. In this context, it is also good to point out that we generally do not believe in a true underlying model that we need to recover. Instead, we search for a model that can describes the data relatively good and is stable.

We applied our method to an empirical data set and compared the results to the results obtained using standard separate regression models for each of the responses. We showed that these standard regression models do not take into account the ordered nature of the predictor variables (as they use dummy variables for the categorical predictors), whereas our model does. When fitting separate models, we need many more parameters than in our reduced rank model and we showed that this leads to overfitting, that is, large and unstable parameter estimates. In contrast, our GMR$^3$ model leads to stable results. Furthermore, our model clearly show which response variables are affected in a similar way by the predictors and our model can take into account the ordered nature of some of the predictors.

We let the regression weights and the loadings to be free parameter estimates. In this sense, the method is exploratory, that is, it finds an underlying structure. In some applications, there might be a priori knowledge about the grouping of response variables or predictor variables. Such knowledge might be incorporated in the future by fixing sets of coefficients to zero, for example, such that a few predictors or responses only pertain to the first, or second dimension. 

Often, not all predictor variables or response variables are available for all observations, that is, we have missing data. In this paper, we did not consider missing data, we assumed the data was completely observed. Future work should consider ways for handling missing data within the proposed framework. 

For numeric reduced rank regression models, biplot representations have been developed \citep{braak1994biplots} and also for logistic reduced rank regression models \citep{derooij2023new}. Future work might consider biplot representations for our generalized mixed reduced rank regression model.  

Other future possibilities include the extension of the model to high dimensional predictors (i.e., large $P$) and high dimensional responses (i.e., large $R$). In that case, penalties such as the lasso \citep{tibshirani1996regression}, ridge \citep{hoerl1970ridge}, or the group lasso \citep{yuan2006model} to penalize the regression weights and/or the loadings. Of course, this requires the development and testing of new algorithms and ways to find optimal penalty parameters.   
A final future extension, is the generalization of our GMR$^3$ method to nested or longitudinal data. We assumed the observations to be independent, but in many applications certain observations might be correlated, for example when observe children in schools, or when we follow participants over time. We need to take the dependencies into account, which can be done by including for example random effects as in multilevel models \citep{snijders2011multilevel, kreft1998introducing}, or by adapting the information criteria \citep{pan2001akaike} and the bootstrap \citep{sherman1997comparison, deen2020clusterbootstrap}. Often, in longitudinal data analysis the interest lies in the development over time for different groups. How to answer such a question when we include optimal scaling of the group variable is something that needs to be investigated. 

\section*{Appendix A: comparison to separate models}

We fitted standard binary and ordinal logistic regression models to the seven response variables. As predictors, we used age, political alignment, gender, degree of urbanization, and education level.
The categorical predictor variables were coded using dummies, with the first category as baseline. As, all predictors were of importance in our gmr3 model, we did not do any further model selection.
The sum of deviances of these seven models is 9923.01 (corresponding to $\mathcal{L}$ of 4961.51). Of course this value is lower compared to our analysis, because more parameters are fitted. That is, the total number of parameters is 115. The corresponding AIC equals 10153.01, while the BIC is 10696.94. Both values are higher than those for our selected model.

The parameter estimates obtained using the seven separate models are given in Table \ref{tab:separatemodels}.

\begin{table}[ht]
\centering
\begin{tabular}{rrrrrrrr}
\hline
& T & FE & CI & MW & FS & DI & RE \\
\hline
A & -0.18 & -0.32 & -0.04 & 0.09 & 0.16 & 0.46 & 0.25 \\
PA2 & -0.96 & -0.36 & -0.81 & -0.59 & -0.56 & 0.17 & -0.59 \\
PA3 & -1.30 & -0.98 & -1.40 & -0.99 & -0.72 & 0.35 & -1.04 \\
G2 & -0.18 & 0.02 & -0.22 & 0.02 & -0.40 & -0.53 & -0.21 \\
U2 & 0.06 & -0.01 & 0.03 & -0.04 & 0.11 & -0.00 & 0.02 \\
U3 & 0.69 & 0.15 & 0.37 & 0.18 & 0.40 & -0.07 & 0.41 \\
E2 & -0.71 & 0.12 & -1.54 & -10.14 & -13.26 & -12.48 & -11.87 \\
E3 & -0.35 & 0.33 & -0.99 & -11.19 & -12.36 & -12.99 & -11.88 \\
E4 & -0.26 & 0.04 & -0.98 & -10.92 & -12.03 & -13.22 & -12.14 \\
E5 & -0.13 & 2.24 & -0.05 & -11.02 & -12.05 & -12.51 & -11.27 \\
E6 & 0.22 & -0.07 & -0.71 & -11.08 & -12.21 & -13.19 & -11.85 \\
E7 & 0.09 & 0.14 & -0.21 & -11.19 & -12.00 & -13.46 & -11.93 \\
E8 & 0.65 & 0.14 & -0.18 & -11.43 & -11.71 & -13.18 & -12.04 \\
E9 & 1.52 & 0.87 & 0.51 & -10.14 & -10.04 & -13.02 & -12.55 \\
\hline
\end{tabular}
\caption{Coefficients of separately fitted logistic or ordinal logistic regression models}
\label{tab:separatemodels}
\end{table}

To compare the fitted parameters of the separate models to those of our model, we need to compute some implied coefficients. The regression coefficients corresponding to the dummy variables, specify a difference in the log odds between two categories of a predictor variable. We can obtain similar coefficients from our model. We already saw the implied coefficients $\bm{BV}'$. To obtain the effect of two predictor categories, we need to incorporate the optimal scaling.
Suppose, we like to know the effect of the second category against the first of the first predictor on the different response variables. Then we need to compute $(\phi_1(2) - \phi_1(1))\bm{b}'_1\bm{V}'$. Here
$\bm{b}_1$ is the $S$-dimensional vector of coefficients for the first predictor, and $\phi_1(1)$ is the optimally scaled value for category 1 of the first predictor variable and $\phi_1(2)$ the optimally scaled value for category 2 of the first predictor variable. Similarly, for the third against the first category of the first predictor we need $(\phi_1(3) - \phi_1(1))\bm{b}'_1\bm{V}'$. In this way, we can compare the effects obtained in the separate models to the effects obtained in our model. The implied coefficients are given in Table \ref{tab:impliedgmr3}

\begin{table}[ht]
\centering
\begin{tabular}{rrrrrrrr}
\hline
& T & FE & CI & MW & FS & DI & RE \\
\hline
A & -0.16 & -0.27 & -0.05 & 0.06 & 0.17 & 0.50 & 0.23 \\
PA2 & -0.96 & -0.42 & -0.90 & -0.33 & -0.57 & 0.16 & -0.46 \\
PA3 & -1.55 & -0.68 & -1.45 & -0.53 & -0.92 & 0.26 & -0.75 \\
G2 & -0.12 & 0.13 & -0.21 & -0.14 & -0.31 & -0.41 & -0.34 \\
U2 & 0.04 & 0.02 & 0.04 & 0.02 & 0.03 & -0.00 & 0.02 \\
U3 & 0.53 & 0.20 & 0.51 & 0.20 & 0.35 & -0.02 & 0.30 \\
E2 & 0.00 & 0.00 & 0.00 & 0.00 & 0.00 & -0.00 & 0.00 \\
E3 & 0.28 & 0.13 & 0.25 & 0.09 & 0.15 & -0.08 & 0.11 \\
E4 & 0.33 & 0.16 & 0.30 & 0.11 & 0.18 & -0.09 & 0.14 \\
E5 & 0.64 & 0.31 & 0.58 & 0.20 & 0.34 & -0.18 & 0.26 \\
E6 & 0.64 & 0.31 & 0.58 & 0.20 & 0.34 & -0.18 & 0.26 \\
E7 & 0.81 & 0.39 & 0.74 & 0.26 & 0.43 & -0.23 & 0.33 \\
E8 & 0.97 & 0.47 & 0.89 & 0.31 & 0.51 & -0.27 & 0.40 \\
E9 & 1.77 & 0.86 & 1.62 & 0.56 & 0.94 & -0.49 & 0.73 \\
\hline
\end{tabular}
\caption{Implied coefficients of GMR3 model}
\label{tab:impliedgmr3}
\end{table}

Whereas some of the coefficients of the separate models are really large, we see that all our coefficients behave well. Because, we assumed some predictor variables to be ordinal, we see that the implied coefficients are neatly ordered, whereas such ordering is absent from the fitted separate models.

We can also look at the standard errors of these coefficients. Applying the bootstrap, and computing these coefficients in each of the bootstrap samples, we can obtain a bootstrap standard error by computing the standard deviation over the bootstrap samples. For the separately fitted models these standard errors are given in Table \ref{tab:bootseparate}. The corresponding standard errors of our model are given in Table \ref{tab:bootimpliedgmr3}. It can be verified that the results of our model are much more stable, that is, the standard errors are much smaller overall. In other words, the efficiency of the our estimator is much higher, specifically for the categorical predictors.

\begin{table}[ht]
\centering
\begin{tabular}{rrrrrrrr}
\hline
& T & FE & CI & MW & FS & DI & RE \\
\hline
A & 0.08 & 0.08 & 0.07 & 0.08 & 0.08 & 0.07 & 0.07 \\
PA2 & 0.20 & 0.18 & 0.16 & 0.19 & 0.20 & 0.16 & 0.16 \\
PA3 & 0.22 & 0.21 & 0.21 & 0.21 & 0.24 & 0.19 & 0.20 \\
G2 & 0.17 & 0.16 & 0.15 & 0.17 & 0.18 & 0.15 & 0.15 \\
U2 & 0.18 & 0.17 & 0.16 & 0.18 & 0.19 & 0.16 & 0.17 \\
U3 & 0.22 & 0.20 & 0.19 & 0.21 & 0.24 & 0.20 & 0.21 \\
E2 & 11.17 & 10.79 & 0.84 & 4.76 & 3.62 & 4.61 & 4.47 \\
E3 & 10.08 & 9.96 & 0.42 & 3.22 & 3.55 & 4.27 & 4.37 \\
E4 & 10.09 & 9.97 & 0.40 & 3.22 & 3.55 & 4.27 & 4.36 \\
E5 & 10.86 & 11.88 & 0.87 & 5.25 & 6.18 & 6.16 & 6.35 \\
E6 & 10.10 & 9.99 & 0.45 & 3.24 & 3.58 & 4.29 & 4.36 \\
E7 & 10.09 & 9.98 & 0.40 & 3.23 & 3.55 & 4.28 & 4.38 \\
E8 & 10.10 & 9.98 & 0.40 & 3.22 & 3.55 & 4.27 & 4.36 \\
E9 & 10.71 & 10.08 & 0.52 & 2.56 & 5.34 & 0.86 & 0.99 \\
\hline
\end{tabular}
\caption{Bootstrap standard error estimates of separately fitted models}
\label{tab:bootseparate}.
\end{table}

\begin{table}[ht]
\centering
\begin{tabular}{rrrrrrrr}
\hline
& T & FE & CI & MW & FS & DI & RE \\
\hline
A & 0.09 & 0.08 & 0.07 & 0.10 & 0.08 & 0.08 & 0.09 \\
PA2 & 0.16 & 0.10 & 0.13 & 0.13 & 0.13 & 0.12 & 0.14 \\
PA3 & 0.20 & 0.17 & 0.19 & 0.20 & 0.17 & 0.20 & 0.21 \\
G2 & 0.14 & 0.11 & 0.13 & 0.10 & 0.14 & 0.15 & 0.11 \\
U2 & 0.10 & 0.04 & 0.09 & 0.04 & 0.07 & 0.04 & 0.06 \\
U3 & 0.17 & 0.11 & 0.15 & 0.11 & 0.15 & 0.16 & 0.15 \\
E2 & 0.37 & 0.21 & 0.31 & 0.09 & 0.16 & 0.22 & 0.12 \\
E3 & 0.52 & 0.30 & 0.43 & 0.13 & 0.24 & 0.32 & 0.17 \\
E4 & 0.54 & 0.32 & 0.45 & 0.14 & 0.26 & 0.35 & 0.20 \\
E5 & 0.54 & 0.33 & 0.45 & 0.15 & 0.26 & 0.38 & 0.21 \\
E6 & 0.54 & 0.33 & 0.45 & 0.16 & 0.27 & 0.39 & 0.22 \\
E7 & 0.54 & 0.35 & 0.45 & 0.17 & 0.28 & 0.44 & 0.24 \\
E8 & 0.64 & 0.39 & 0.53 & 0.22 & 0.37 & 0.50 & 0.31 \\
E9 & 0.70 & 0.48 & 0.57 & 0.31 & 0.46 & 0.64 & 0.40 \\
\hline
\end{tabular}
\caption{Bootstrap standard error estimates of GMR3 models}
\label{tab:bootimpliedgmr3}
\end{table}







\bibliographystyle{apalike}
\bibliography{gmr4.bib}  

\begin{thebibliography}{}

\bibitem[Agresti, 2013]{agresti2013categorical}
Agresti, A. (2013).
\newblock {\em Categorical data analysis}.
\newblock John Wiley \& Sons, third edition.

\bibitem[Anderson and Philips, 1981]{anderson1981regression}
Anderson, J. and Philips, P. (1981).
\newblock Regression, discrimination and measurement models for ordered
  categorical variables.
\newblock {\em Journal of the Royal Statistical Society: Series C (Applied
  Statistics)}, 30(1):22--31.

\bibitem[Anderson, 1951]{anderson1951estimating}
Anderson, T.~W. (1951).
\newblock Estimating linear restrictions on regression coefficients for
  multivariate normal distributions.
\newblock {\em The Annals of Mathematical Statistics}, pages 327--351.

\bibitem[Busing, 2022]{busing2022monotone}
Busing, F.~M. (2022).
\newblock Monotone regression: A simple and fast o (n) pava implementation.
\newblock {\em Journal of Statistical Software}, 102:1--25.

\bibitem[Davies and Tso, 1982]{davies1982procedures}
Davies, P. and Tso, M. K.-S. (1982).
\newblock Procedures for reduced-rank regression.
\newblock {\em Journal of the Royal Statistical Society: Series C (Applied
  Statistics)}, 31(3):244--255.

\bibitem[Davison and Hinkley, 1997]{davison1997bootstrap}
Davison, A.~C. and Hinkley, D.~V. (1997).
\newblock {\em Bootstrap methods and their application}.
\newblock Cambridge university press.

\bibitem[De~Leeuw, 2005]{deleeuw2005monotonic}
De~Leeuw, J. (2005).
\newblock Monotonic regression.
\newblock {\em Encyclopedia of Statistics in Behavioral Science}.

\bibitem[De~Leeuw, 2006]{deleeuw2006principal}
De~Leeuw, J. (2006).
\newblock Principal component analysis of binary data by iterated singular
  value decomposition.
\newblock {\em Computational Statistics \& Data Analysis}, 50(1):21--39.

\bibitem[De~Rooij, 2023]{derooij2023new}
De~Rooij, M. (2023).
\newblock A new algorithm and a discussion about visualization for logistic
  reduced rank regression.
\newblock {\em Behaviormetrika}, pages 1--22.

\bibitem[De~Rooij et~al., 2023]{derooijbreemerwoestenburgbusing2022}
De~Rooij, M., Breemer, L., Woestenburg, D., and Busing, F. M. T.~A. (2023).
\newblock Logistic multidimensional data analysis for ordinal response
  variables using a cumulative link function.
\newblock {\em Submitted paper}.

\bibitem[Deen and de~Rooij, 2020]{deen2020clusterbootstrap}
Deen, M. and de~Rooij, M. (2020).
\newblock Clusterbootstrap: An r-package for the analysis of hierarchical data
  using generalized linear models with the cluster bootstrap.
\newblock {\em Behavior Research Methods}, 52:572--590.

\bibitem[Efron, 1979]{efron1979bootstrap}
Efron, B. (1979).
\newblock Bootstrap methods: Another look at the jackknife.
\newblock {\em Annals of Statistics}, 7:1--26.

\bibitem[Efron and Tibshirani, 1986]{efron1986bootstrap}
Efron, B. and Tibshirani, R. (1986).
\newblock Bootstrap methods for standard errors, confidence intervals, and
  other measures of statistical accuracy.
\newblock {\em Statistical Science}, 1:54--75.

\bibitem[European~Commission, 2023]{ZA7953}
European~Commission, B. (2023).
\newblock Eurobarometer 98.2 (2023).
\newblock GESIS, Cologne. ZA7953 Data file Version 1.0.0,
  https://doi.org/10.4232/1.14081.

\bibitem[Fish, 1988]{fish1988multivariate}
Fish, L.~J. (1988).
\newblock Why multivariate methods are usually vital.
\newblock {\em Measurement and Evaluation in Counseling and Development},
  21(3):130--137.

\bibitem[Friendly et~al., 2013]{friendly2013elliptical}
Friendly, M., Monette, G., and Fox, J. (2013).
\newblock Elliptical insights: understanding statistical methods through
  elliptical geometry.
\newblock {\em Statistical Science}, pages 1--39.

\bibitem[Gifi, 1990]{gifi1990nonlinear}
Gifi, A. (1990).
\newblock {\em Nonlinear multivariate analysis}.
\newblock Wiley-Blackwell.

\bibitem[Hastie et~al., 2009]{hastie2009elements}
Hastie, T., Tibshirani, R., and Friedman, J. (2009).
\newblock {\em The elements of statistical learning}.
\newblock Springer series in statistics New York.

\bibitem[Heiser, 1995]{heiser1995convergent}
Heiser, W.~J. (1995).
\newblock Convergent computation by iterative majorization: Theory and
  applications in multidimensional data analysis.
\newblock In Krzanowski, W.~J., editor, {\em Recent Advances in Descriptive
  Multivariate Analysis}, pages 157--189. Clarendon Press.

\bibitem[Hoerl and Kennard, 1970]{hoerl1970ridge}
Hoerl, A.~E. and Kennard, R.~W. (1970).
\newblock Ridge regression: Biased estimation for nonorthogonal problems.
\newblock {\em Technometrics}, 12(1):55--67.

\bibitem[Hunter and Lange, 2004]{hunter2004tutorial}
Hunter, D.~R. and Lange, K. (2004).
\newblock A tutorial on {MM} algorithms.
\newblock {\em The American Statistician}, 58(1):30--37.

\bibitem[Izenman, 1975]{izenman1975reduced}
Izenman, A.~J. (1975).
\newblock Reduced-rank regression for the multivariate linear model.
\newblock {\em Journal of Multivariate Analysis}, 5(2):248--264.

\bibitem[Jiao, 2016]{jiao2016high}
Jiao, F. (2016).
\newblock {\em High-dimensional inference of ordinal data with medical
  applications}.
\newblock PhD thesis, University of Iowa.

\bibitem[Kreft and De~Leeuw, 1998]{kreft1998introducing}
Kreft, I.~G. and De~Leeuw, J. (1998).
\newblock {\em Introducing multilevel modeling}.
\newblock Sage.

\bibitem[Luo et~al., 2018]{luo2018leveraging}
Luo, C., Liang, J., Li, G., Wang, F., Zhang, C., Dey, D.~K., and Chen, K.
  (2018).
\newblock Leveraging mixed and incomplete outcomes via reduced-rank modeling.
\newblock {\em Journal of Multivariate Analysis}, 167:378--394.

\bibitem[Mc~Fadden, 1974]{mcfadden1974}
Mc~Fadden, D. (1974).
\newblock Conditional logit analysis of qualitative choice behavior.
\newblock In {\em In P. Zarembka, editor, Frontiers in Econometrics, chapter
  Four}, pages 102--142. Academic Press.

\bibitem[McCullagh, 1980]{mccullagh1980regression}
McCullagh, P. (1980).
\newblock Regression models for ordinal data.
\newblock {\em Journal of the Royal Statistical Society: Series B
  (Methodological)}, 42(2):109--127.

\bibitem[Meulman et~al., 2019]{meulman2019ros}
Meulman, J.~J., van~der Kooij, A.~J., and Duisters, K.~L. (2019).
\newblock Ros regression: Integrating regularization with optimal scaling
  regression.
\newblock {\em Statistical Science}, 34(3):361--390.

\bibitem[Nguyen, 2017]{nguyen2017introduction}
Nguyen, H.~D. (2017).
\newblock An introduction to majorization-minimization algorithms for machine
  learning and statistical estimation.
\newblock {\em Wiley Interdisciplinary Reviews: Data Mining and Knowledge
  Discovery}, 7(2):e1198.

\bibitem[Pan, 2001]{pan2001akaike}
Pan, W. (2001).
\newblock Akaike's information criterion in generalized estimating equations.
\newblock {\em Biometrics}, 57(1):120--125.

\bibitem[Sherman and Cessie, 1997]{sherman1997comparison}
Sherman, M. and Cessie, S.~l. (1997).
\newblock A comparison between bootstrap methods and generalized estimating
  equations for correlated outcomes in generalized linear models.
\newblock {\em Communications in Statistics-Simulation and Computation},
  26(3):901--925.

\bibitem[Snijders and Bosker, 2011]{snijders2011multilevel}
Snijders, T.~A. and Bosker, R. (2011).
\newblock {\em Multilevel analysis: An introduction to basic and advanced
  multilevel modeling}.
\newblock sage.

\bibitem[Song et~al., 2021]{song2021generalized}
Song, Y., Westerhuis, J.~A., Aben, N., Wessels, L. F.~A., Groenen, P. J.~F.,
  and Smilde, A.~K. (2021).
\newblock Generalized simultaneous component analysis of binary and
  quantitative data.
\newblock {\em Journal of Chemometrics}, 35(3):e3312.

\bibitem[Stevens, 1946]{stevens1946theory}
Stevens, S.~S. (1946).
\newblock On the theory of scales of measurement.
\newblock {\em Science}, 103(2684):677--680.

\bibitem[Takane, 2013]{takane2013constrained}
Takane, Y. (2013).
\newblock {\em Constrained principal component analysis and related
  techniques}.
\newblock CRC Press.

\bibitem[Ten~Berge, 1993]{tenberge1993}
Ten~Berge, J.~M. (1993).
\newblock {\em Least squares optimization in multivariate analysis}.
\newblock DSWO Press, Leiden University Leiden.

\bibitem[{Ter Braak} and Looman, 1994]{braak1994biplots}
{Ter Braak}, C.~J. and Looman, C.~W. (1994).
\newblock Biplots in reduced-rank regression.
\newblock {\em Biometrical Journal}, 36(8):983--1003.

\bibitem[Tibshirani, 1996]{tibshirani1996regression}
Tibshirani, R. (1996).
\newblock Regression shrinkage and selection via the lasso.
\newblock {\em Journal of the Royal Statistical Society: Series B
  (Methodological)}, 58(1):267--288.

\bibitem[Tso, 1981]{tso1981reduced}
Tso, M.-S. (1981).
\newblock Reduced-rank regression and canonical analysis.
\newblock {\em Journal of the Royal Statistical Society: Series B
  (Methodological)}, 43(2):183--189.

\bibitem[Van~den Wollenberg, 1977]{wollenberg1977redundancy}
Van~den Wollenberg, A.~L. (1977).
\newblock Redundancy analysis an alternative for canonical correlation
  analysis.
\newblock {\em Psychometrika}, 42(2):207--219.

\bibitem[Vapnik, 1994]{vapnik1995}
Vapnik, V. (1994).
\newblock {\em The Nature of Statistical Learning Theory}.
\newblock New York: Springer.

\bibitem[Vapnik, 1998]{vapnik1998}
Vapnik, V. (1998).
\newblock {\em Statistical Learning Theory}.
\newblock New York: Wiley.

\bibitem[Willems, 2020]{willems2020advances}
Willems, S. (2020).
\newblock {\em Advances in Survival Analysis and Optimal Scaling Methods}.
\newblock PhD thesis, Leiden University.

\bibitem[Yee, 2015]{yee2015}
Yee, T.~W. (2015).
\newblock {\em Vector Generalized Linear and Additive Models: With an
  Implementation in R}.
\newblock Springer, New York, USA.

\bibitem[Yee and Hastie, 2003]{yee2003reduced}
Yee, T.~W. and Hastie, T.~J. (2003).
\newblock Reduced-rank vector generalized linear models.
\newblock {\em Statistical Modelling}, 3(1):15--41.

\bibitem[Young et~al., 1976]{young1976regression}
Young, F.~W., De~Leeuw, J., and Takane, Y. (1976).
\newblock Regression with qualitative and quantitative variables: An
  alternating least squares method with optimal scaling features.
\newblock {\em Psychometrika}, 41(4):505--529.

\bibitem[Yuan and Lin, 2006]{yuan2006model}
Yuan, M. and Lin, Y. (2006).
\newblock Model selection and estimation in regression with grouped variables.
\newblock {\em Journal of the Royal Statistical Society Series B: Statistical
  Methodology}, 68(1):49--67.

\end{thebibliography}
\end{document}